\documentclass[reprint,aps,prb,showpacs]{revtex4-1}
\usepackage{graphicx}
\usepackage{amsmath} 
\usepackage{amsfonts}
\usepackage{amssymb}
\newcommand{\EqLabel}[1]{\label{#1}}

\begin{document}

\title{Non-equilibrium BBGKY Hierarchy from the Redfield Equation}
\author{Jinshan Wu}
\affiliation{Department of Physics $\&$ Astronomy, University of British Columbia,
\\ Vancouver, B.C. Canada, V6T 1Z1}

\begin{abstract}
A BBGKY-like hierarchy is derived from the non-equilibrium Redfield equation. Two further approximations are introduced and each can be used to truncate and solve the hierarchy. In the first approximation such a truncation is performed by replacing two-particle Green's functions (GFs) in the hierarchy by their values at equilibrium. The second method is developed based on the cluster expansion, which constructs two-particle GFs from one-particle GFs and neglects the correlation part. A non-equilibrium Wick's Theorem is proved to provide a basis for this non-equilibrium cluster expansion. Using those two approximations, our method of solving the Redfield equation, for instance, of an $N$-site chain of interacting spinless fermions, involves an eigenvalue problem with dimension $2^{N}$ and a linear system with dimension $N^2$ in the first case, and a nonlinear equation with dimension $N^2$ in the second case, which can be solved iteratively via a sequence of $N^2$ linear systems. Other currently available direct methods correspond to a linear system or an eigenvalue system with dimension $4^N$ plus an eigenvalue system with dimension $2^N$. As a test of the methods, for small systems with size $N=4$, results are found to be consistent with results made available by other direct methods. Although not discussed here, extending both methods to their next levels is straightforward. This indicates a promising potential for this BBGKY-like approach of non-equilibrium kinetic equations.
\end{abstract}
\maketitle

\section{Introduction}

A dynamic equation, such as Newton's equation or the Schr\"{o}dinger's equation, describes only dynamical processes and strictly speaking it does not describe evolution towards thermal equilibrium, although sometimes it is used to do so together with a presumed thermal equilibrium distribution, such as in the derivation of Kubo formula of linear response theory\cite{Kubo}. Starting from the dynamic equation and its corresponding BBGKY hierarchy\cite{BBGKY} the uniqueness of a stationary solution at thermal equilibrium, and particularly being the Boltzmann distribution, remains as an assumption but not a proved theorem\cite{BBGKY}. The Redfield equation\cite{Redfield0}, although it may show an unphysical transient process due to the use of the Markov approximation, on the other hand puts the description of dynamical evolution, thermal evolution towards the equilibrium state and evolution towards non-equilibrium stationary states (NESSs) under a common framework. 

The Redfield equation has long been a standard tool in the study of the relaxation process in the theory of nuclear magnetic resonance\cite{Redfield1}, optical spectroscopy and chemical dynamical systems\cite{Redfield2}. Due to the fact that usually in those studies the central system of interest is modeled by a Hamiltonian with a very low dimension, the lack of an efficient algorithm to solve the Redfield equation, other than direct diagonalization and direct integration, is not a serious problem. However, when the Redfield equation is applied to transport calculations, the size of the system is usually much larger. It is then necessary to have a more efficient way to find the NESSs, without which many physical questions remain unanswered.

A famous example is the validity of the phenomenology law of transport. Advancement in science and technology has made it possible to build nanoscale electronic devices\cite{MesoReview}, where classical phenomenological laws of transport such as Fourier's Law and Ohm's Law may not be valid any more\cite{MesoReview}. In order to construct a theory of transport for mesoscopic or microscopic systems, and also to check under what circumstances those phenomenological laws hold, one may start from the first principles, i.e. the dynamic equation of classical or quantum systems, and add in as few extra assumptions as possible. The Landauer formula\cite{Landauer} makes use of scattering waves from the Schr\"{o}dinger's equation but a biased distribution of those waves is assumed to calculate physical quantities. The non-equilibrium Green's function (NEGF) method starts from two decoupled systems each at their own thermal equilibrium states, which could be very far from the expected non-equilibrium stationary states (NESSs), and treats coupling between the two systems perturbatively\cite{NEGF}. For interacting systems, usually NEGF is used together with the density functional method\cite{Guo}, from which the whole spectrum and corresponding effective wavefunctions are calculated to construct the non-equilibrium density matrix. Both the perturbation and the effective wavefunctions introduce further approximations to the calculation. 

The Redfield equation approach to studying transport phenomena is to explicitly couple the system of interest, $H_{S}$, to reservoirs and then use the projector technique\cite{KuboBook} to derive an effective equation of motion for the system. One then solves this Redfield equation to get NESSs. The Redfield equation requires validity of the Markovian approximation and treats the coupling between system and reservoirs within second-order perturbation. This generalization of the Redfield equation to transport studies was first implemented by Saito\cite{Saito_TwoBath} and more or less followed by others\cite{Michel, Jinshan, Li}. In such cases one quite often expects to deal with systems with size $N\approx 100$ in order to be comparable to other methods such as the Landauer formula and NEGF\cite{MesoReview}. This then leads to a Hilbert space with dimension $2^{100}$, when we for example consider an $N$-site chain of spinless fermions with $N=100$, meaning a matrix with dimension $4^{100}$ in the corresponding Redfield equation. Due to this exponential increasing of the problem size, and the lack of a more efficient alternative method, currently one can only discuss physical behaviors of very small systems with $N\approx10$\cite{Jinshan, Li}. Conclusions drawn from numerical results for such small systems are not regarded as reliable enough. In this work, we will present a new approach, using the idea of a BBGKY hierarchy to study this kinetic equation. This allows us to develop very efficient and systematic approximate methods to find NESSs. 

A general Redfield equation can be cast into the following form\cite{Jinshan},
\begin{widetext}
\begin{align}
\frac{\partial \rho\left(t\right)}{\partial t} = L_{H_{S}}\rho\left(t\right) + L_{V_{in}}\rho\left(t\right) + \lambda^2 L_{B}\left(T, \mu\right)\rho\left(t\right) + \lambda^2 L_{P}\left(\Delta T, \Delta \mu\right)\rho\left(t\right) \equiv L\rho\left(t\right), 
\EqLabel{EOM} 
\end{align}
\end{widetext}
where $L_{H_{S}}\rho = -i\left[H_{S}, \rho\right]$ is the Hamiltonian of the central system. Sometimes we also use $H_{S}=H_{0}+V$ to separate $H_{S}$ into non-interacting part $H_{0}$ and interaction $V$. Finally $L_{V_{in}}\rho = -i\left[V_{in}, \rho\right]$, where $V_{in}$ is a possible induced potential, for example electric potential due to charge distribution in the case of charge transport. $L_{B}\left(T, \mu\right)$ comes from coupling to baths with coupling strength $\lambda$, and $L_{P}\left(\Delta T, \delta \mu\right)$ exists when baths have different temperatures and/or chemical potentials. This equation describes a dynamical process when $\lambda=0$, thermal relaxation towards equilibrium when $\lambda\neq0,\Delta T=0, \Delta \mu=0$, and evolution towards NESSs when they are all nonzero. 

If we are interested in the long-time steady-state solution $\rho_{\infty}$, then
\begin{align}
L\rho_{\infty}=0,
\EqLabel{NESSEOM}
\end{align} 
which is sometimes called a stationary Redfield equation. Here $L$ is a matrix with dimension $d^2$, where $d$ is the dimension of the systems Hilbert space, for example, $d=2^N$ for the $N$-site chain of spinless fermions mentioned above. Solving this equation is very costly computationally. It involves solving a linear system\cite{Jinshan} or an eigenvalue system\cite{Michel} with dimension $4^N$ plus an eigenvalue system with dimension $2^N$. Currently, for interacting central systems, one usually can only solve the Redfield equation numerically up to $N=10$\cite{Li, Michel, Jinshan} while for non-interacting systems, the equation with $N$ around a hundred can be solved in terms of single-particle Green's functions (GFs)\cite{Saito_TwoBath}. For example, for non-interacting systems, in Ref.\onlinecite{Saito_TwoBath}, a closed equation of single-particle GFs, $G_{1}\left(k^{\dag}, k^{'}\right) \equiv \langle c^{\dag}_{k}c_{k^{'}}\rangle =tr\left(c^{\dag}_{k}c_{k^{'}}\rho_{\infty}\right)$, was derived from the stationary Redfield equation Eq\eqref{NESSEOM} and solved. 

Our idea is basically to extend this GF based solution of the Redfield equation for non-interacting systems onto interacting systems. We consider $G_{1}\left(m^{\dag}, m^{'}\right)$, GFs in the lattice basis, and derive an equation for these GFs. In the presence of interaction, the single-particle GFs $G_{1}\left(m^{\dag}, m^{'}\right)$, generally denoted as $G_{1}$, will be coupled to the two-particle GFs $G_{2}\left(m^{\dag}, n^{\dag}, m^{'}, n^{'}\right) \equiv \langle c^{\dag}_{m}c^{\dag}_{n}c_{m^{'}}c_{n^{'}}\rangle $, also generally denoted as $G_{2}$, which are then coupled to three-particle GFs, $G_{3}\left(l^{\dag}, m^{\dag}, n^{\dag}, l^{'}, m^{'}, n^{'}\right) \equiv \langle c^{\dag}_{l}c^{\dag}_{m}c^{\dag}_{n}c_{l^{'}}c_{m^{'}}c_{n^{'}}\rangle $, also generally denoted as $G_{3}$ and so on. We arrive at a BBGKY-like equation hierarchy\cite{BBGKY}. 

In principle, solving the whole hierarchy is as hard as solving directly the Redfield equation. The hierarchy has to be truncated first and then solved. In the rest part of this paper, after deriving the hierarchy from the Redfield equation, we will then present two such methods. In the example calculations, we will only truncate the hierarchy at the first equation of the hierarchy. The first method substitutes value of $G_{2}$ at equilibrium for the unknown $G_{2}$ appearing in the first equation while the second expresses $G_{2}$ as combinations of $G_{1}$ via cluster expansion. After either one, the equation is closed and then solved. We will see in the following that both methods are significantly more efficient than the direct methods. The first one is capable of dealing with relatively small systems but with large interaction strength while the second one can deal with much larger systems but with relatively small interaction strength.    

\section{Derivation of BBGKY-like Hierarchy}

For concreteness in presenting our general formulation, let us start from a Redfield equation describing an $N$-site chain of spinless fermions coupled with two fermionic baths. Our system of interest is defined by $H_{S}$,
\begin{widetext}
\begin{align}
\EqLabel{HS}
H_{S}=-t\sum_{l=1}^{N-1}\left(c^{\dag}_{l}c_{l+1}+c^{\dag}_{l+1}c_{l}\right) + V_{0}\sum_{l=1}^{N-1}c^{\dag}_{l+1}c_{l+1}c^{\dag}_{l}c_{l} = H_{0}+V.
\end{align}
\end{widetext}
The two heat baths are collections of fermionic modes,
\begin{align}
{\cal H}_B = \sum_{k, \alpha} \omega_{k,\alpha}
b^{\dag}_{k,\alpha}b_{k,\alpha},
\end{align}
where $\alpha=L, R$ indexes the left and right-side baths and we set
$\hbar=1, k_B=1$, the lattice constant $a=1$ and hopping constant $t=1$. The system-baths coupling
is chosen as:
\begin{align}
V = \lambda \sum_{k,\alpha} V^{\alpha}_{k}\left(c^{\dag}_{\alpha}b_{k,\alpha}+c_{\alpha}b^{\dag}_{k,\alpha}\right),
\end{align}
where the left (right) bath is coupled to the first (last) site: $c_{L}=c_{1}$ and $c_{R}=c_{N}$ and so on. Bath parameters, including temperature and chemical potential, are chosen to be $\left(T_{L}, \mu\right)$ and $\left(T_{R}, \mu\right)$ with $T_{L/R}=T\pm\frac{\Delta T}{2}$. In the present work, the induced $L_{V_{in}}$ term in Eq\eqref{EOM}, is neglected.

The corresponding Redfield equation reads\cite{KuboBook, Jinshan},
\begin{widetext}
\begin{align}
\EqLabel{evol} \frac{\partial \rho(t)}{\partial t} =-i[{\cal
H}_S, \rho(t)]-\lambda^2\sum_{\alpha=L,R}^{}\left\{
\left[c^{\dag}_{\alpha}, \hat{m}_{\alpha} \rho(t)\right] + \left[c_{\alpha}, \hat{\bar{m}}_{\alpha} \rho(t)\right] +
h.c.\right\},
\end{align}
\end{widetext}
where $\hat{m}_{L}$($\hat{m}_{R}$) is related to $c_{1}$($c_{N}$) and $\hat{\bar{m}}_{L}$($\hat{\bar{m}}_{R}$) is related to $c^{\dag}_{1}$($c^{\dag}_{N}$)\cite{KuboBook},
\begin{subequations}
\EqLabel{ms}
\begin{align}
\hat{m}_{\alpha} = \sum_{k}|V^{\alpha}_{k}|^2\int_{0}^{\infty}d\tau c_{\alpha}\left(-\tau\right) e^{-i\omega_{k,\alpha}\tau} \langle 1-n\left(\omega_{k,\alpha}\right)\rangle  ,
\\ \hat{\bar{m}}_{\alpha} = \sum_{k}|V^{\alpha}_{k}|^2\int_{0}^{\infty}d\tau c^{\dag}_{\alpha}\left(-\tau\right) e^{i\omega_{k,\alpha}\tau} \langle n\left(\omega_{k,\alpha}\right)\rangle  . 
\end{align}
\end{subequations}
Here $n\left(\omega_{k,\alpha}\right)=\left(e^{\beta_{\alpha}\left(\omega_{k, \alpha}-\mu_{\alpha}\right)}+1\right)^{-1}$ is the Fermi-Dirac distribution with the bath temperature $T_\alpha=1/\beta_\alpha$ and chemical potential $\mu_{\alpha}$. If $U\left(t\right)=e^{-iH_{S}t}$ is known, then so is $c_{\alpha}\left(t\right)=U^{\dag}\left(t\right)c_{\alpha}U\left(t\right)$ and therefore the operators $\hat{m}$. This requires a full diagonalization of $H_{S}$. Using eigenvectors of $H_{S}$, one can perform the above integrals to get operator $\hat{m}$s. Detail is included in Appendix \ref{sec:mPerturbation}. There we will also see that change of variable between summation over $k$ and integration over energy in Eq\eqref{ms} involves the density of states of the baths $D_\alpha\left(\omega\right)$. We combine this density of states together with coupling constant $V^{\alpha}_{k}$, and set $D_\alpha\left(\Omega_{mn}\right)|V^{\alpha}_{k_{mn}}|^2$ (see Appendix \ref{sec:mPerturbation} for detail) as an overall constant, which is included in $\lambda^2$.   

When only a long-time steady-state solution $\rho_{\infty}$ is of interest, we may derive a stationary form, Eq\eqref{NESSEOM}, from the kinetic Redfield equation. Furthermore, for a physical quantity of the central system with operator $A$, from Eq\eqref{NESSEOM}, we have generally
\begin{widetext}
\begin{align}
\EqLabel{eq:A}
0=i\langle \left[A, H_{0}\right]\rangle  + i\langle \left[A, V\right]\rangle  
+ \lambda^2\sum_{\alpha}\left\{\langle \left[A, c^{\dag}_{\alpha}\right]\hat{m}_{\alpha}\rangle  + \langle \left[A, c_{\alpha}\right]\hat{\bar{m}}_{\alpha}\rangle  - \langle \hat{m}^{\dag}_{\alpha}\left[A, c_{\alpha}\right]\rangle  - \langle \hat{\bar{m}}^{\dag}_{\alpha}\left[A, c^{\dag}_{\alpha}\right]\rangle \right\},
\end{align}
\end{widetext}
where $\hat{m}^{\dag}_{\alpha}$($\hat{\bar{m}}^{\dag}_{\alpha}$) is the hermitian conjugate of $\hat{m}_{\alpha}$($\hat{\bar{m}}_{\alpha}$). All equations of GFs in the rest of this paper will be derived from this equation. For example the first and the second equation of the hierarchy can be derived from using $A=c^{\dag}_{m}c_{n}$ and $A=c^{\dag}_{m}c^{\dag}_{n}c_{m^{'}}c_{n^{'}}$ in Eq\eqref{eq:A},
\begin{widetext}
\begin{subequations}
\EqLabel{A:G1G2}
\protect
\begin{align}
\EqLabel{A1:g1}
0= it\langle c^{\dag}_{m-1}c_{n}\rangle  + it\langle c^{\dag}_{m+1}c_{n}\rangle  -it\langle c^{\dag}_{m}c_{n+1}\rangle   -it\langle c^{\dag}_{m}c_{n-1}\rangle  \\
\EqLabel{A1:g2}
 - iV_{0}\langle c^{\dag}_{m}c^{\dag}_{n-1}c_{n}c_{n-1}\rangle  + iV_{0}\langle c^{\dag}_{m+1}c^{\dag}_{m}c_{m+1}c_{n}\rangle  
 - iV_{0}\langle c^{\dag}_{n+1}c^{\dag}_{m}c_{n+1}c_{n}\rangle  + iV_{0}\langle c^{\dag}_{m}c^{\dag}_{m-1}c_{n}c_{m-1}\rangle   \\
\EqLabel{A1:gm}
-\lambda^2\sum_{\alpha}\langle \delta_{m\alpha}c_{n}\hat{\bar{m}}_{\alpha} + \delta_{n\alpha}\hat{\bar{m}}^{\dag}_{\alpha}c^{\dag}_{m} - \delta_{n\alpha}c^{\dag}_{m}\hat{m}_{\alpha} - \delta_{m\alpha}\hat{m}^{\dag}_{\alpha}c_{n}\rangle , 
\end{align}  
\end{subequations}
and
\begin{subequations}
\EqLabel{A:G2G3}
\protect
\begin{align}
\EqLabel{A2:g2}
0= it\langle c^{\dag}_{m}c^{\dag}_{n}c_{m^{'}}c_{n^{'}+1}\rangle  + it\langle c^{\dag}_{m}c^{\dag}_{n}c_{m^{'}}c_{n^{'}-1}\rangle  + it\langle c^{\dag}_{m}c^{\dag}_{n}c_{m^{'}+1}c_{n^{'}}\rangle  + it\langle c^{\dag}_{m}c^{\dag}_{n}c_{m^{'}-1}c_{n^{'}}\rangle  \notag \\
- it\langle c^{\dag}_{m}c^{\dag}_{n-1}c_{m^{'}}c_{n^{'}}\rangle  - it\langle c^{\dag}_{m}c^{\dag}_{n+1}c_{m^{'}}c_{n^{'}}\rangle  - it\langle c^{\dag}_{m-1}c^{\dag}_{n}c_{m^{'}}c_{n^{'}}\rangle   - it\langle c^{\dag}_{m+1}c^{\dag}_{n}c_{m^{'}}c_{n^{'}}\rangle  \notag \\
+ iV_{0}\langle c^{\dag}_{m}c^{\dag}_{n}c_{m^{'}}c_{n^{'}}\rangle \left(\delta_{m^{'}+1, n^{'}}+\delta_{m^{'}-1, n^{'}}-\delta_{m+1, n}-\delta_{m-1, n}\right) \\
\EqLabel{A2:g3}
 - iV_{0}\sum_{l=m\pm1, n\pm1}\langle c^{\dag}_{l}c^{\dag}_{m}c^{\dag}_{n}c_{l}c_{m^{'}}c_{n^{'}}\rangle  + iV_{0}\sum_{l=m^{'}\pm1, n^{'}\pm1}\langle c^{\dag}_{l}c^{\dag}_{m}c^{\dag}_{n}c_{l}c_{m^{'}}c_{n^{'}}\rangle  \\
\EqLabel{A2:gm}
-\lambda^2\sum_{\alpha}\langle \delta_{m^{'}\alpha}c^{\dag}_{m}c^{\dag}_{n}c_{n^{'}}\hat{m}_{\alpha} - \delta_{n^{'}\alpha}c^{\dag}_{m}c^{\dag}_{n}c_{m^{'}}\hat{m}_{\alpha} + \delta_{m\alpha}c^{\dag}_{n}c_{m^{'}}c_{n^{'}}\hat{\bar{m}}_{\alpha} - \delta_{n\alpha}c^{\dag}_{m}c_{m^{'}}c_{n^{'}}\hat{\bar{m}}_{\alpha}\rangle  \notag \\
-\lambda^2\sum_{\alpha}\langle \delta_{n^{'}\alpha}\hat{\bar{m}}^{\dag}_{\alpha}c^{\dag}_{m}c^{\dag}_{n}c_{m^{'}} -\delta_{m^{'}\alpha}\hat{\bar{m}}^{\dag}_{\alpha}c^{\dag}_{m}c^{\dag}_{n}c_{n^{'}} + \delta_{n\alpha}\hat{m}^{\dag}_{\alpha}c^{\dag}_{m}c_{m^{'}}c_{n^{'}} - \delta_{m\alpha}\hat{m}^{\dag}_{\alpha}c^{\dag}_{n}c_{m^{'}}c_{n^{'}} \rangle .
\end{align}  
\end{subequations}
\end{widetext}
Note that since the set of all polynomials of $\left\{c_{l}, c^{\dag}_{l}\right\}$ forms a complete basis of the operator space, operators $\hat{m}$ are certain functions of polynomials of $\left\{c_{l}, c^{\dag}_{l}\right\}$. Therefore, as expected $G_{1}$ is coupled to $G_{2}$ from Eq\eqref{A1:g2}, and possibly also $G_{3}$ or higher GFs from Eq\eqref{A1:gm}, and $G_{2}$ is coupled to $G_{3}$ from Eq\eqref{A2:g3}, and possibly also $G_{4}$ or higher GFs from Eq\eqref{A2:gm}. Solving such an equation hierarchy is no easier than directly solving the Redfield equation, unless $V_{0}=0$ so that the above equation of $G_{1}$ is closed and is not coupled to $G_{2}$. 

We may, however, solve these equations by truncating the hierarchy at certain order with some further approximations, such as the molecular-chaos assumption in the classical Boltzmann equation\cite{KuboBook}, or replacing high order GFs by cluster expansion of lower order ones\cite{Koch, Kadanoff}. In this work, we suggest the following two approximate methods: $(1)$ substitution of certain high-order GFs by their values at equilibrium; $(2)$ expressing high-order GFs as combinations of lower-order ones plus a correlation part via cluster expansion and then ignoring the correlation part at certain order. Specifically in the following example calculation, the first-order form of both approximations, i.e. only the first equation of the hierarchy is used and substitution or cluster expansion is preformed on $G_{2}$. One can do such a substitution or cluster expansion of GFs at further-order GFs and make use of further equations in the hierarchy. A general discussion of accuracy of such substitutions at different orders will be presented elsewhere. In this work, we focus on the potential of this BBGKY-like formulation and discuss briefly the topic of performance of the two approximations in their first-order forms.

\section{Solving the Hierarchy}
In order to solve Eq\eqref{A:G1G2} explicitly, we will first have to find explicit forms of operators $\hat{m}$ in terms of operator $\left\{c_{l}, c^{\dag}_{l}\right\}$. In the following we will present one exact numerical calculation and one perturbative calculation of those operators. Correspondingly based on these two methods of finding operators $\hat{m}$, we will discuss in this section two ways of making Eq\eqref{A:G1G2} to be a closed equation by dealing with $G_{2}$ terms in the equation differently. 

We will first discuss a more accurate even for a large $V_{0}$ but computationally costly method, perturbation based on two-particle GFs at equilibrium. Next we will discuss a relatively less accurate but computationally much cheaper method, the non-equilibrium cluster expansion. The later works only for relatively small $V_{0}$ but it can be applied on much larger systems. The unknown $G_{1}\left(m^{\dag},n\right)$ solved from both methods will be compared against, $G^{Ex}_{1}\left(m^{\dag},n\right)$, the exact solution of Eq\eqref{NESSEOM}. A measure of relative distance between two matrices $A$ and $B$,
\begin{align}
\EqLabel{deq}
d^{A}_{B}=\frac{\sqrt{\sum_{ij} \left|A_{ij}-B_{ij}\right|^2}}{\sqrt{\sum_{ij} \left|B_{ij}\right|^2}},
\end{align} 
is used to describe the accuracy of our approximations.

\subsection{Method $1$: Starting from Equilibrium States}
As explicitly worked out in Eq\eqref{Exms} in Appendix \ref{sec:mPerturbation}, operator $\hat{m}$s can be written in eigenmodes of $H_{S}$, which can be solved from an exact diagonalization of $H_{S}$, a $2^N$-dimension eigenvalue problem. Then in the language of super-operator space\cite{Michel}, where operators are treated like vectors -- so called super-vectors, super-vectors $\hat{m}$ can be expanded under the basis --- polynomials of $\left\{c_{l}, c^{\dag}_{l}\right\}$,   
\begin{subequations}
\EqLabel{mdecomposition1}
\begin{align}
\hat{m}_{\alpha} = \sum_{l} d_{\alpha; l}c_{l}+V_{0}D_{\alpha} \\
\hat{\bar{m}}_{\alpha} = \sum_{l} \bar{d}_{\alpha; l}c^{\dag}_{l}+V_{0}\bar{D}_{\alpha}.
\end{align}
\end{subequations}
Here we keeps only the linear polynomial in the present work although further expansion is possible. Using the definition of inner product between super-vectors $\langle \langle A|B\rangle \rangle =tr\left(A^{\dag}B\right)$, we have
\begin{subequations}
\begin{align}
d_{\alpha; l}=\frac{1}{2^{\left(N-1\right)}}tr\left(c^{\dag}_{l}\hat{m}_{\alpha}\right), \\
\bar{d}_{\alpha; l}=\frac{1}{2^{\left(N-1\right)}}tr\left(c_{l}\hat{\bar{m}}_{\alpha}\right),
\end{align}
\end{subequations}
and operators $V_{0}D$ and $V_{0}\bar{D}$ are just the rest part of operators $\hat{m}$ and $\hat{\bar{m}}$ respectively. Here $2^{N-1}$ is a normalization constant to make $d_{\alpha, l}=1$ when $\hat{m}_{\alpha}=c_{l}$. 

With above expressions of operators $\hat{m}$, Eq\eqref{A:G1G2} becomes,
\begin{widetext}
\begin{subequations}
\EqLabel{Eq:G1G2}
\protect
\begin{align}
\EqLabel{Eq:g1}
0= it\langle c^{\dag}_{m-1}c_{n}\rangle  + it\langle c^{\dag}_{m+1}c_{n}\rangle  -it\langle c^{\dag}_{m}c_{n+1}\rangle   -it\langle c^{\dag}_{m}c_{n-1}\rangle  \notag \\
+\lambda^2\sum_{l, \alpha}\langle \delta_{n\alpha}\left(d_{\alpha; l} + \bar{d}^{*}_{\alpha; l}\right)c^{\dag}_{m}c_{l} + \delta_{m\alpha}\left(\bar{d}_{\alpha; l} + d^{*}_{\alpha; l}\right)c^{\dag}_{l}c_{n}\rangle   \\
\EqLabel{Eq:nu}
-\lambda^2\sum_{\alpha}\left[\delta_{m\alpha}\bar{d}_{\alpha; n} + \delta_{n\alpha}\bar{d}^{*}_{\alpha; m}\right] \\
\EqLabel{Eq:g2}
 - iV_{0}\langle c^{\dag}_{m}c^{\dag}_{n-1}c_{n}c_{n-1}\rangle  + iV_{0}\langle c^{\dag}_{m+1}c^{\dag}_{m}c_{m+1}c_{n}\rangle  
 - iV_{0}\langle c^{\dag}_{n+1}c^{\dag}_{m}c_{n+1}c_{n}\rangle  + iV_{0}\langle c^{\dag}_{m}c^{\dag}_{m-1}c_{n}c_{m-1}\rangle   \\
\EqLabel{Eq:gd2}
-\lambda^2V_{0}\sum_{\alpha}\langle \delta_{m\alpha}c_{n}\bar{D}_{\alpha} + \delta_{n\alpha}\bar{D}^{\dag}_{\alpha}c^{\dag}_{m} - \delta_{n\alpha}c^{\dag}_{m}D_{\alpha} - \delta_{m\alpha}D^{\dag}_{\alpha}c_{n}\rangle . 
\end{align}  
\end{subequations}
\end{widetext} 
Notice that every $c_{0}, c^{\dag}_{0}, c_{N+1}$ and $c^{\dag}_{N+1}$ that appears in the equation should be recognized as $0$. First let us replace all $G_{2}$s in Eq\eqref{Eq:g2} by their values at equilibrium, denoted here as $G^{E, (0)}_{2}$ where the superscript $(0)$ means to use the thermal equilibrium (denoted by superscript $E$) as the zeroth order approximation of the non-equilibrium $G_{2}$. Using the first term as an example,
\begin{align}
G^{E, (0)}_{2}\left(m^{\dag}, n\right) = tr\left(c^{\dag}_mc^{\dag}_{n-1}c_{n}c_{n-1}\rho_{eq}\left(H_{S}\right)\right),
\end{align}
where $\rho_{eq}\left(H_{S}\right) = \frac{1}{Z}e^{-\frac{H_{S}}{T}}$.
This requires eigenstates of $H_{S}$. Similarly one can define $G^{E,(0)}_{D}$ from Eq\eqref{Eq:gd2}, using the first term as an example, 
\begin{align}
G^{E, (0)}_{D}\left(m^{\dag}, n\right) = \delta_{m\alpha}tr\left(c_n\bar{D}_{\alpha}\rho_{eq}\left(H_{S}\right)\right).
\end{align} 
Next let us calculate $G^{E, (1)}_{1}$ from Eq\eqref{Eq:G1G2}, where the superscript $(1)$ means that the approximate calculation takes care of the first equation of the hierarchy, Eq\eqref{Eq:G1G2}. We organize all $G^{E, (1)}_{1}\left(m^{\dag}, n\right)$ as a vector, 
\begin{align}
g^{E, (1)}_{1}=\left[G_{1}\left(1^{\dag}, 1\right), G_{1}\left(1^{\dag}, 2\right), \cdots, G_{1}\left(N^{\dag}, N\right)\right]^{T},
\end{align}
then Eq\eqref{Eq:G1G2} for given value of $m,n$ is the equation occupying the $\left(mN+n\right)$th row and totally there are $N^2$ such equations. After substituting $G^{E, (0)}_{2}$ and $G^{E,(0)}_{D}$ for the exact but unknown $G_{2}$ and $G_{D}$, the whole set of Eq\eqref{Eq:G1G2} for all $m,n$ then becomes a linear system on $g^{E, (1)}_{1}$ with dimension $N^2$,
\begin{align}
\Gamma^{(1)} g^{E, (1)}_{1}=iV_{0}g_{2}^{E, (0)}+\lambda^2\nu + \lambda^2V_{0}g^{E, (0)}_{D},
\EqLabel{Eq:Gammag1}
\end{align}
where vector $\nu$ comes from ordering Eq\eqref{Eq:nu} in the same way as $g^{E, (1)}_{1}$. The same holds for $g^{E,(0)}_{2}$ and $g^{E,(0)}_{D}$ correspondingly from ordering Eq\eqref{Eq:g2} and Eq\eqref{Eq:gd2}, and from Eq\eqref{Eq:g1} one gets matrix $\Gamma^{(1)}$. For example, assuming $m,n$ are not at boundaries, one may read from Eq\eqref{Eq:G1G2}, 
\begin{subequations}
\begin{align}
\nu_{mN+n}=\sum_{\alpha}\left[\delta_{m\alpha}\bar{d}_{\alpha; n} + \delta_{n\alpha}\bar{d}^{*}_{\alpha; m}\right], \\
\Gamma^{(1)}_{\left(mN+n\right), \left(\left(m-1\right)N+n\right)}=it.
\end{align}
\end{subequations}
Next we calculate single-particle equilibrium GFs, $G^{E, (0)}_{1}$, and organize it in the same way into a vector denoted as $g^{E, (0)}_{1}$. In order to set a reference of the accuracy, we compare $d^{E, (0)}$, the difference between the exact solution $g^{Ex}_{1}$ and the zeroth order $g^{E, (0)}_{1}$, and $d^{E, (1)}$, the difference between the exact solution $g^{Ex}_{1}$ and the first order solution above, $g^{E, (1)}_{1}$. Here $G^{Ex}\left(m^{\dag},n\right)=tr\left(c^{\dag}_{m}c_{n}\rho_{\infty}\right)$, where $\rho_{\infty}$ is the exact solution from Eq\eqref{NESSEOM}. 
 
\subsubsection{results}
\label{sec:result1}
\begin{figure}
\includegraphics[width=\columnwidth]{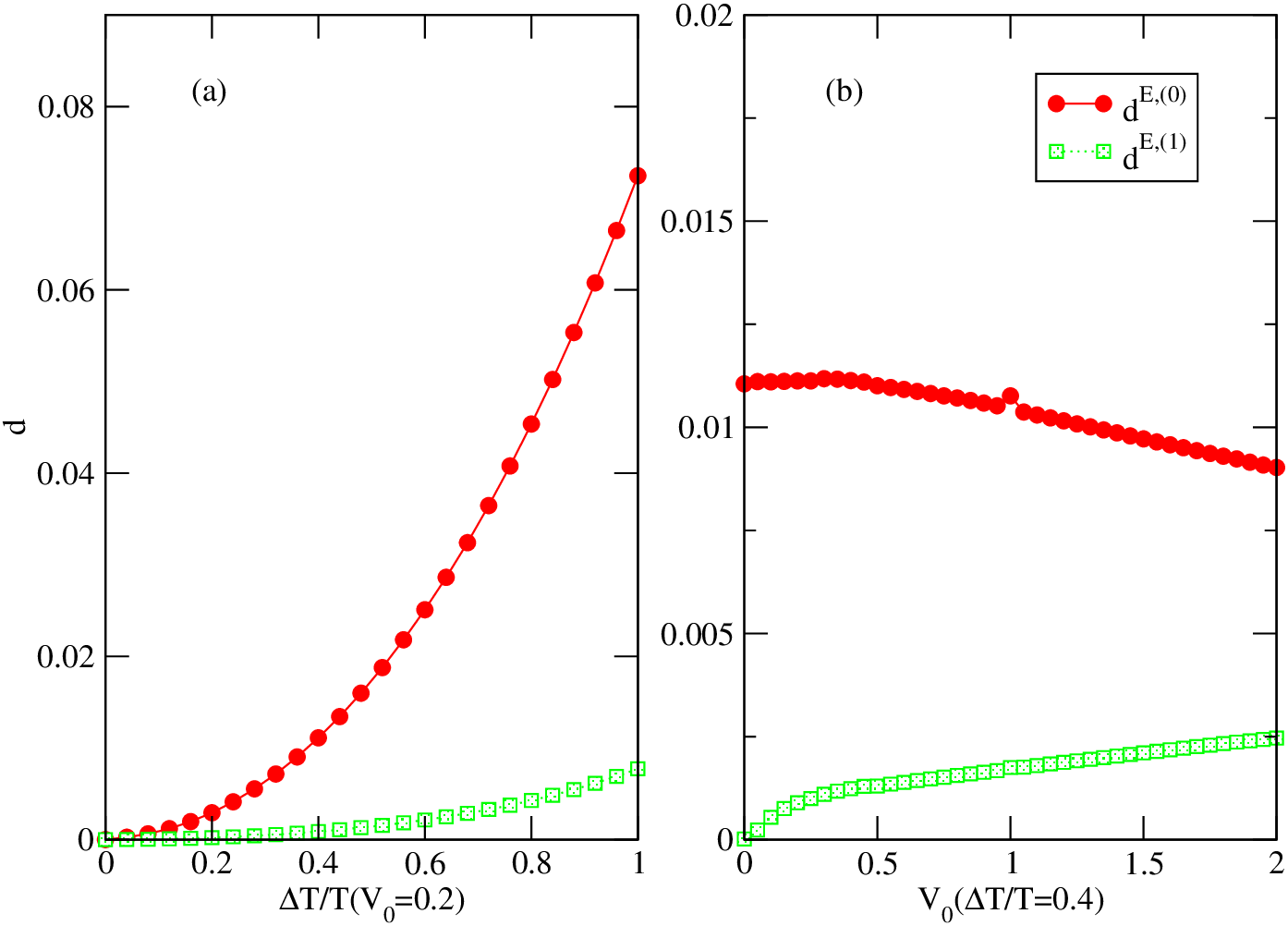} 
\includegraphics[width=\columnwidth]{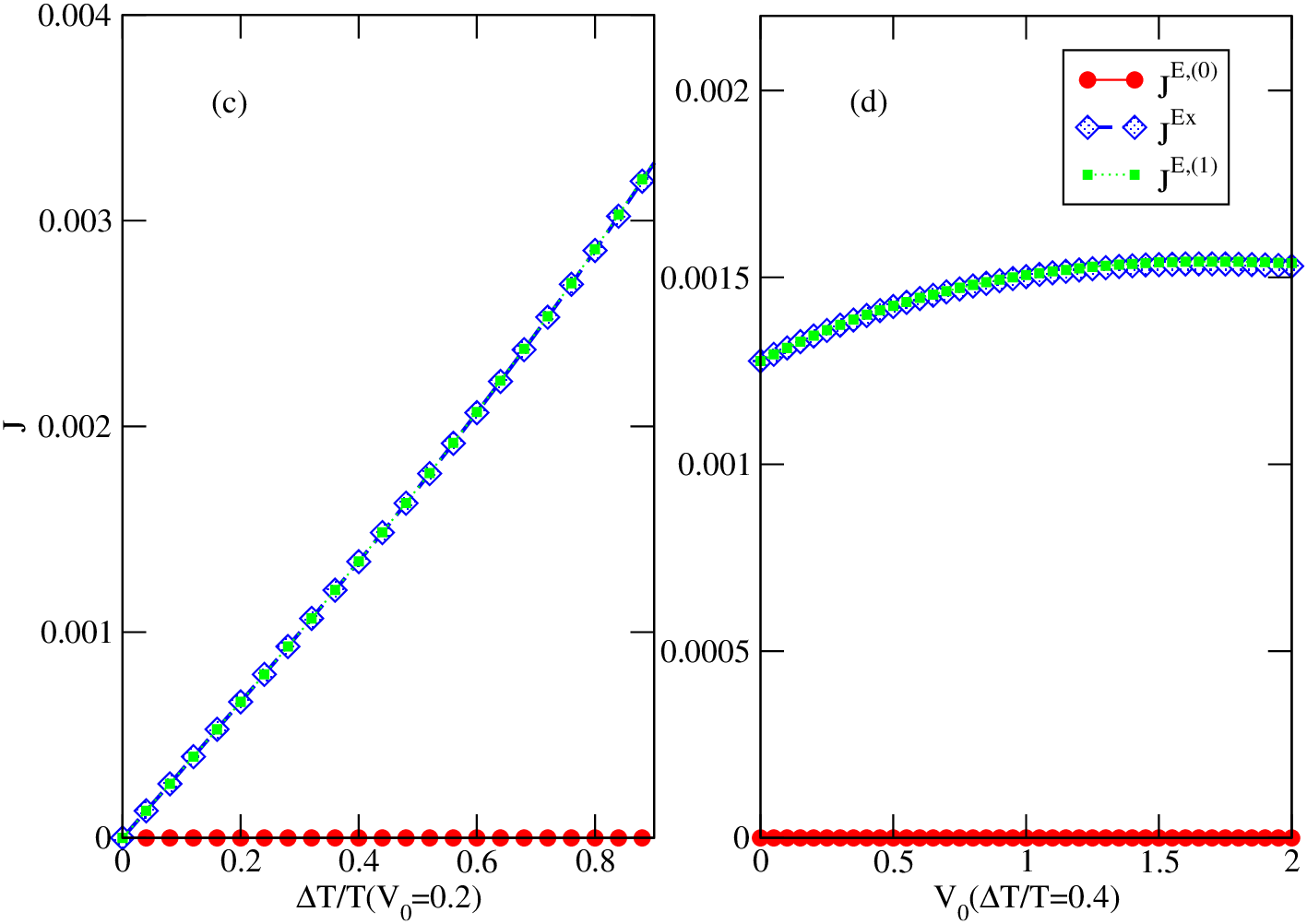}
\caption{\label{fig:Eq} $g^{E, (1)}_{1}$ is compared against $g^{Ex}_{1}$ for interacting systems at non-equilibrium. (a) $V=0.2$, $d^{E,(1)}$is compared with the reference $d^{E,(0)}$ for different values of $\Delta T$. With larger $\Delta T$, $d^{(1)}$ becomes larger but still much smaller than and $d^{E,(0)}$. (b) $\Delta T=0.4T$, accuracy was checked for different values of $V_{0}$. At the worst cases shown in the plot, $d^{(1)}$ is about $0.3\%$, where $V_{0}=2t$ is a relatively large strength of interaction. Electrical currents $J^{E,(0)}$, $J^{E,(1)}$ are compared against $J^{Ex}$ in (c) and (d). We see that $J^{E,(0)}$ is zero while $J^{E,(1)}$ is close to $J^{Ex}$ for even relatively large $V_{0}$. In all these example calculations, $t=1.0$, $\lambda=0.1$, $\mu=-1.0$. }
\end{figure}

First, we set $V_{0}=0.2$ as a constant, and check the accuracy of $g^{E, (1)}_{1}$ with different values of $\Delta T$. From Fig.\ref{fig:Eq}(a) we can see that the worst case is about $d^{(1)}=1\%$. Secondly, we set $\Delta T=0.4T$ as a constant, and check the accuracy of $g^{E, (1)}_{1}$ with different values of $V_{0}$. The worst case is $d^{(1)}=0.3\%$ as shown in Fig.\ref{fig:Eq}(b). Overall, $d^{E, (1)}$ is always much smaller than $d^{E, (0)}$. 
$J^{E,(0)}$, $J^{E,(1)}$, $J^{Ex}$ are calculated respectively from $g_{1}^{E,(0)}$, $g_{1}^{E,(1)}$ and $g_{1}^{Ex}$. From Fig.\ref{fig:Eq}(c) and (d) we see that in both cases, $J^{E,(1)}$ is very close to the exact one $J^{Ex}$ while $J^{E, (0)}$ current in the equilibrium state, is always zero. Very high accuracy is found especially for small $\Delta T$. This indicates that the approximation captures the essential part of the non-equilibrium stationary states. It is also worth mentioning that this method generates reasonable results for very large $V_{0}$. Furthermore, it is likely the approximation could be improved: from further expansions in terms of higher order polynomials of $c_{l}, c^{\dag}_{l}$ and substituting their values at equilibrium for the higher order unknown GFs in higher-order equations of the hierarchy. Stopping the expansion of operators $\hat{m}$ at linear order of $V_{0}$ is compatible with the solving only the first equation of the hierarchy. If further equations of the hierarchy are used then one should also expand operators $\hat{m}$ in further orders of $V_{0}$.      

In order to estimate the accuracy of the first-order form of this approximation and also to get an overview of accuracy of possibly the next order, let us study the leading order of residues in terms of $\lambda^2$ and $\frac{\Delta T}{T}$, which are assumed to be small in the following. Hence $\lambda^2V_{0} \ll V_{0}$, therefore we know that $g_{D}$ is relatively smaller than the other $g_{2}$ term so we drop it. This in fact requires $\lambda^2V_{0} \ll T$, which we assume to be true. Similarly for the same reason since $\lambda^2\Delta T \ll \Delta T$, we drop $\lambda^2\Delta T$ term in $\lambda^2\nu$ in Eq\eqref{Eq:Gammag1},
\begin{align}
\lambda^2\nu = \lambda^2\nu_{0}\left(T\right) + \lambda^2\Delta T \nu_{, T},
\end{align}
and keep only the major term, $\lambda^2\nu_{0}\left(T\right)$, which is independent of $\Delta T$. Here $\nu_{, T}$ denotes formally a derivative of $T$ on $\nu$. The general idea is then to write down respectively equations for $g^{Ex}_{1}$ and $g^{E, (1)}_{1}$, and then compare the two equations to estimate $\Delta^{E,(1)}_{1}=g^{E, (1)}_{1}-g^{Ex}_{1}$. In order to get some information on how such approximation at the next order improves the accuracy, we also want to compare $\Delta^{E,(1)}_{1}$ to $\Delta^{E,(0)}_{1}=g^{E, (0)}_{1}-g^{Ex}_{1}$, which is estimated in the same way from the difference between the equations respectively for for $g^{Ex}_{1}$ and $g^{E, (0)}_{1}$. See Appendix \ref{sec:estimation} for detail of those equations and the estimation. Here we summarize the results that 
\begin{align}
\Delta^{E, (0)}_{1} =  \Delta T \left(\Gamma^{(1)}_{0}\right)^{-1}\Gamma^{(1)}_{, T} g^{Ex}_{1} + iV_{0}\left(\Gamma^{(1)}_{0}\right)^{-1}\Delta^{E, (0)}_{2} 
\end{align}
and 
\begin{widetext}
\begin{align}
\Delta^{E, (1)}_{1} =  \left(\Gamma^{(1)}_{0}\right)^{-1}\left[-V^{2}_{0}\left(\Gamma^{(2)}_{0}\right)^{-1}\Delta^{E, (0)}_{3} + iV_{0}\Delta T\left(\Gamma^{(2)}_{0}\right)^{-1}\Gamma^{(2)}_{, T} g^{Ex}_{2} + iV_{0}\lambda^2\left(\Gamma^{(2)}_{0}\right)^{-1}\Delta^{E, (0)}_{1}\right].
\EqLabel{Eq:DeltaFinal}
\end{align}
\end{widetext}
Here $\Delta^{E, (0)}_{n}=g^{E, (0)}_{n}-g^{Ex}_{n}$ and $\Delta^{E, (1)}_{n}=g^{E, (1)}_{n}-g^{Ex}_{n}$. We refer readers to Appendix \ref{sec:estimation} for definitions of all $\Gamma$ matrices. Most importantly here we see that $\Delta^{E, (0)}_{1}$ is multiplied by a small number $\lambda^2V_{0}$ and then becomes a part of $\Delta^{E, (1)}_{1}$. Furthermore, this relation holds generally for higher-order forms of this approximation. Judged from this term, as long as $\lambda^2V_{0}$ is a small number compared with $t$ then the method under consideration is very reasonable. As of the other two additional terms, they can be regarded as $\left(V^{2}_{0}g^{Ex}_{3}+V_{0}g^{Ex}_{2}\right)\Delta T$. Therefore, the limit of $\frac{V_{0}}{t}$ where this method is still applicable is by $\left|g^{Ex}_{2}\right|^{-1}$ or $\left|g^{Ex}_{3}\right|^{-\frac{1}{2}}$, which could be much larger than $1$ since roughly $\left|g^{Ex}_{n}\right|=\left|g^{Ex}\right|^{n}$ --- the smaller the larger $n$. This explains why as we see from Fig.\ref{fig:Eq} that this method is applicable even for $V_{0}$ larger than $t$. We have also tested several systems with larger $N$ (up to $N=8$) and no qualitative difference on accuracy has been found. More detail and more systematic analysis will be presented elsewhere.

\subsection{Method $2$: Non-equilibrium Cluster Expansion}

Another way to make Eq\eqref{A:G1G2} to be a closed equation is to use cluster expansion. In the case of equilibrium GFs it proposes for example at the level of two-particle GFs,
\begin{widetext}
\begin{subequations}
\begin{align}
G_{2}\left(m^{\dag},n^{\dag},m^{'},n^{'}\right) = -G_{1}\left(m^{\dag}, m^{'}\right)G_{1}\left(n^{\dag},n^{'}\right) 
+ G_{1}\left(m^{\dag},n^{'}\right)G_{1}\left(n^{\dag},m^{'}\right) + \mathfrak{G}_{2}\left(m^{\dag}, n^{\dag}, m^{'}, n^{'}\right),
\end{align}
\end{subequations}
\end{widetext}
and then sets
\begin{align}
\mathfrak{G}_{2}=0.
\end{align}
It can be applied on higher-order GFs, for example by a similar expansion on $G_{3}$ and setting $\mathfrak{G}_{3}=0$. The fact that equilibrium Wick's Theorem shows that indeed $\mathfrak{G}_{2}=0$ when $V_{0}=0$ makes this expansion plausible for equilibrium GFs. Here in the case of non-equilibrium GFs, we are going to propose the same expansion and this requires a non-equilibrium Wick's Theorem to be true that $\mathfrak{G}_{2}=0$ when $V_{0}=0$ holds for non-equilibrium GFs. Fortunately, this can be proved (see Appendix \ref{sec:wick} for detail). Setting $\mathfrak{G}_{2}=0$ with $V_{0}\neq 0$ is like using the Hartree-Fock approximation, so that depending on the system and the physical problem under investigation, one may quite often need to go beyond that to the next level of approximation, i.e. keeping $\mathfrak{G}_{2}$ but ignoring $\mathfrak{G}_{3}$ and truncating the equation hierarchy at the second equation instead of the first equation. In this work, we will use the first level approximation, i.e. ignoring $\mathfrak{G}_{2}$. 

However, with operators $D$ defined in the previous section, cluster expansion could not be applied, we have to expand operators operators $\hat{m}$ in higher-order polynomials of $\left\{c_{l}, c^{\dag}_{l}\right\}$. This can be done as following.       
Other than the exact direct diagonalization, operators $\hat{m}$ can also be found perturbatively analytically(see Appendix \ref{sec:mPerturbation} for the full detail). The basic idea is to start from assuming 
\begin{align}
c_{l}\left(t\right)=c^{(0)}_{l}\left(t\right) + V_{0}c^{(1)}_{l}\left(t\right) + O\left(V^{2}_{0}\right),
\end{align}
and then derive and solve the equations of motion of $c^{(0)}_{l}, c^{(1)}_{l}$ from the Heisenberg's equation.
In this way, one avoids the direct diagonalization of $H_{S}$ so that it simplifies the calculation but its accuracy depends on the order of $V_{0}$ at which the expansion stops. Stopping at the linear order of $V_{0}$ is compatible with the cluster expansion at $G_{2}$. If cluster expansion at higher-order GFs is applied then operators $\hat{m}$ should also be expanded in higher orders of $V_{0}$. Kept only the first order, operators $\hat{m}$ become
\begin{widetext} 
\begin{subequations}
\EqLabel{mdecomposition2}
\begin{align}
\hat{m}_{\alpha} = \sum_{m} \mathfrak{D}_{\alpha; m}c_{m}+V_{0}\sum_{m_{1}m_{2}m_{3}}\mathfrak{D}_{\alpha; m_{1}m_{2}m_{3}}c_{m_{1}}c^{\dag}_{m_{2}}c_{m_{3}} +O\left(V^2_{0}\right) \\
\hat{\bar{m}}_{\alpha} = \sum_{m} \bar{\mathfrak{D}}_{\alpha; m}c^{\dag}_{m}-V_{0}\sum_{m_{1}m_{2}m_{3}}\mathfrak{D}_{\alpha; m_{1}m_{2}m_{3}}c^{\dag}_{m_{3}}c_{m_{2}}c^{\dag}_{m_{1}}  +O\left(V^2_{0}\right),
\end{align}
\end{subequations}
\end{widetext}  
where definitions of $\mathfrak{D}_{\alpha; m}$ and $\mathfrak{D}_{\alpha; m_{1}m_{2}m_{3}}$ are given in Appendix \ref{sec:mPerturbation}.

With the first-order cluster expansion and the above expansion of operators $\hat{m}$ plugged into Eq\eqref{A:G1G2}, we have
\begin{widetext}
\begin{subequations}
\EqLabel{G1G2cluster}
\protect
\begin{align}
\EqLabel{cluster:g1}
0= itG_{1}\left(m-1;n\right) + itG_{1}\left(m+1;n\right) - itG_{1}\left(m;n+1\right) - itG_{1}\left(m;n-1\right) \notag \\
+\lambda^2\sum_{l, \alpha}\left[\delta_{n\alpha}\left(\mathfrak{D}_{\alpha; l} + \bar{\mathfrak{D}}^{*}_{\alpha; l}\right)G_{1}\left(m;l\right) + \delta_{m\alpha}\left(\bar{\mathfrak{D}}_{\alpha; l} + \mathfrak{D}^{*}_{\alpha; l}\right)G_{1}\left(l;n\right)\right] \\
+\lambda^2V_{0}\sum_{\alpha, m_{1}, m_{2}}\left(\mathfrak{D}_{\alpha;nm_{2}m_{1}}-\mathfrak{D}_{\alpha;m_{1}m_{2}n}\right)G_{1}\left(m_{1},m_{2}\right)\delta_{m\alpha} \notag \\
+\lambda^2V_{0}\sum_{\alpha, m_{1}, m_{2}}\left(\mathfrak{D}_{\alpha;mm_{2}m_{1}}-\mathfrak{D}_{\alpha;m_{1}m_{2}m}\right)G_{1}\left(m_{2},m_{1}\right)\delta_{n\alpha} \\
\EqLabel{cluster:nu0}
-\lambda^2\sum_{\alpha}\left(\delta_{m\alpha}\bar{\mathfrak{D}}_{\alpha; n} + \delta_{n\alpha}\bar{\mathfrak{D}}^{*}_{\alpha; m}\right) \\
\EqLabel{cluster:nu1} 
+ \lambda^2V_{0}\sum_{\alpha, m_{1}}\left(\mathfrak{D}_{\alpha;m_{1}m_{1}n}\delta_{m\alpha} + \mathfrak{D}_{\alpha;m_{1}m_{1}m}\delta_{n\alpha}\right)\\
\EqLabel{cluster:g2}
 - iV_{0}G_{1}\left(m;n-1\right)G_{1}\left(n-1;n\right) + iV_{0}G_{1}\left(m;n\right)G_{1}\left(n-1;n-1\right) \notag \\
 - iV_{0}G_{1}\left(m+1;m+1\right)G_{1}\left(m;n\right) + iV_{0}G_{1}\left(m+1;n\right)G_{1}\left(m;m+1\right) \notag \\
 - iV_{0}G_{1}\left(n+1;n\right)G_{1}\left(m;n+1\right) + iV_{0}G_{1}\left(n+1;n+1\right)G_{1}\left(m;n\right) \notag \\
 - iV_{0}G_{1}\left(m;n\right)G_{1}\left(m-1;m-1\right) + iV_{0}G_{1}\left(m;m-1\right)G_{1}\left(m-1;n\right).
\end{align}  
\end{subequations}
\end{widetext} 
Next we define a vector $g^{C, (1)}_{1}$, where superscript $C$ means cluster expansion and $(1)$ means keeping only the first equation in the hierarchy, similarly as $g^{E, (1)}_{1}$. For simplicity of expressions let us order Eq\eqref{cluster:g2} in the same way and denote it as $g^{C, (1)}_{2}=\Pi\left(g^{C, (1)}_{1}\right)$, where $\Pi$ refers to the nonlinear function --- summation of product --- of $g^{C, (1)}_{1}$ in Eq\eqref{cluster:g2}. Then the above equation can be denoted as
\begin{widetext}
\begin{align}
\EqLabel{cluster:gc1}
\left(\Gamma^{(1)}_{0}+\lambda^2V_{0}\Gamma^{(1)}_{D}\right) g^{C, (1)}_{1} = \lambda^2\nu_{0} + \lambda^2V_{0}\nu_{1} + iV_{0}g^{C, (1)}_{2},
\end{align}  
\end{widetext}
where the five terms are respectively the five sub equations in Eq\eqref{G1G2cluster}, for example,
\begin{align}
\left(\nu_{0}\right)_{mN+n}=\sum_{\alpha}\left(\delta_{m\alpha}\bar{\mathfrak{D}}_{\alpha; n} + \delta_{n\alpha}\bar{\mathfrak{D}}^{*}_{\alpha; m}\right).
\end{align}
This equation can be solve iteratively 
\begin{widetext}
\begin{align}
g^{(n+1)}_{1} = \left(\Gamma^{(1)}_{0}+\lambda^2V_{0}\Gamma^{(1)}_{D}\right)^{-1}\left(\lambda^2\nu_{0} + \lambda^2V_{0}\nu_{1} + iV_{0}\Pi\left(g^{(n)}_{1}\right)\right), g^{(0)}_{1} = \left(\Gamma^{(1)}_{0}\right)^{-1}\lambda^2\nu_{0},
\end{align}  
\end{widetext}
where we start from $g^{(0)}_{1} = g^{C, (0)}_{1}$, which is the exact solution of Eq\eqref{cluster:gc1} when $V_{0}=0$ and through the iteration defined above we get solution $g^{C, (1)}_{1}=\lim_{n\rightarrow \infty}g^{(n)}_{1}$, which in practice stops at large enough $n$ such that $g^{(n)}_{1}-g^{(n-1)}_{1}$ is small enough.

\subsubsection{results}
\label{sec:result2}
\begin{figure}
\includegraphics[width=\columnwidth]{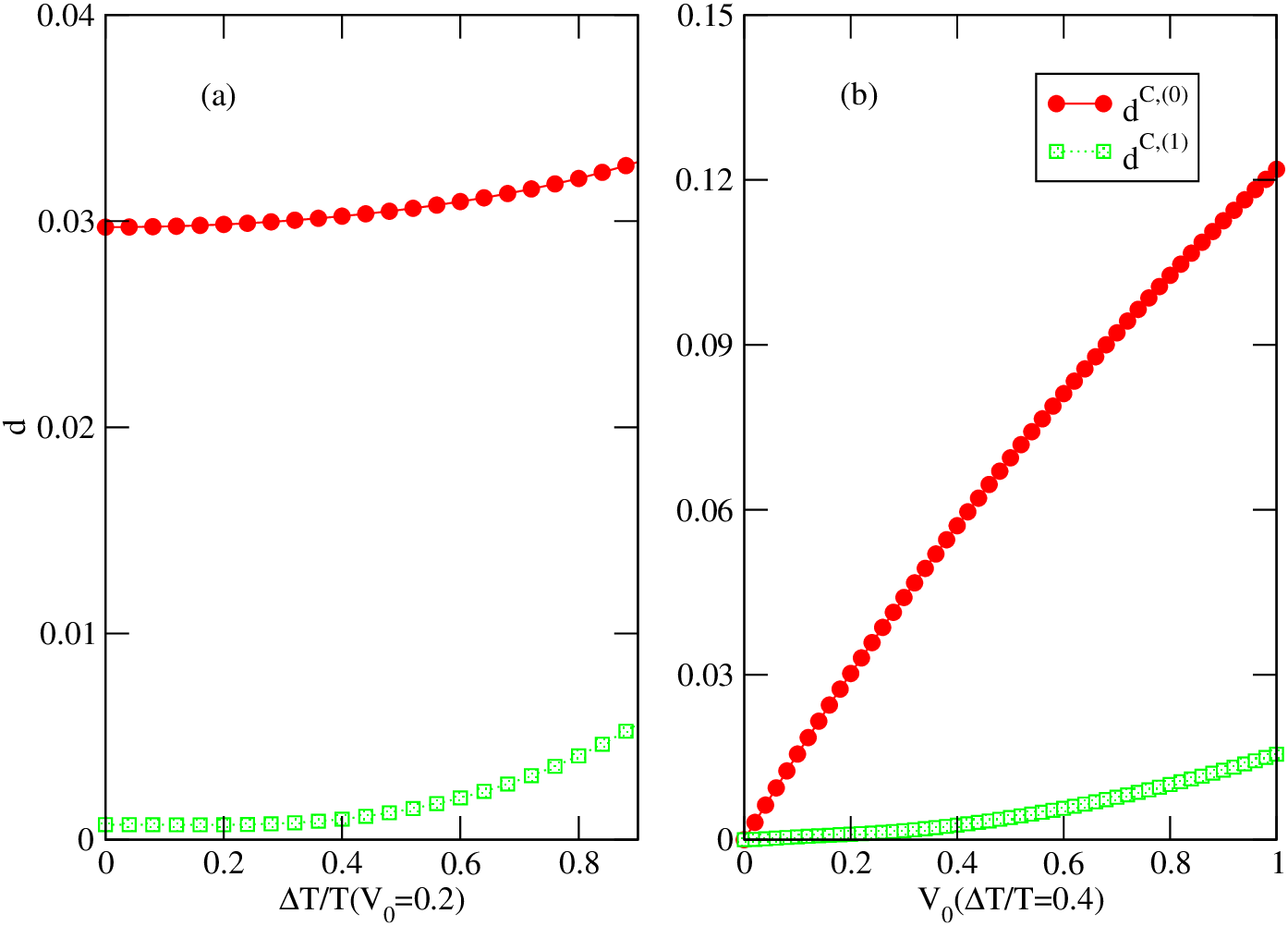} 
\includegraphics[width=\columnwidth]{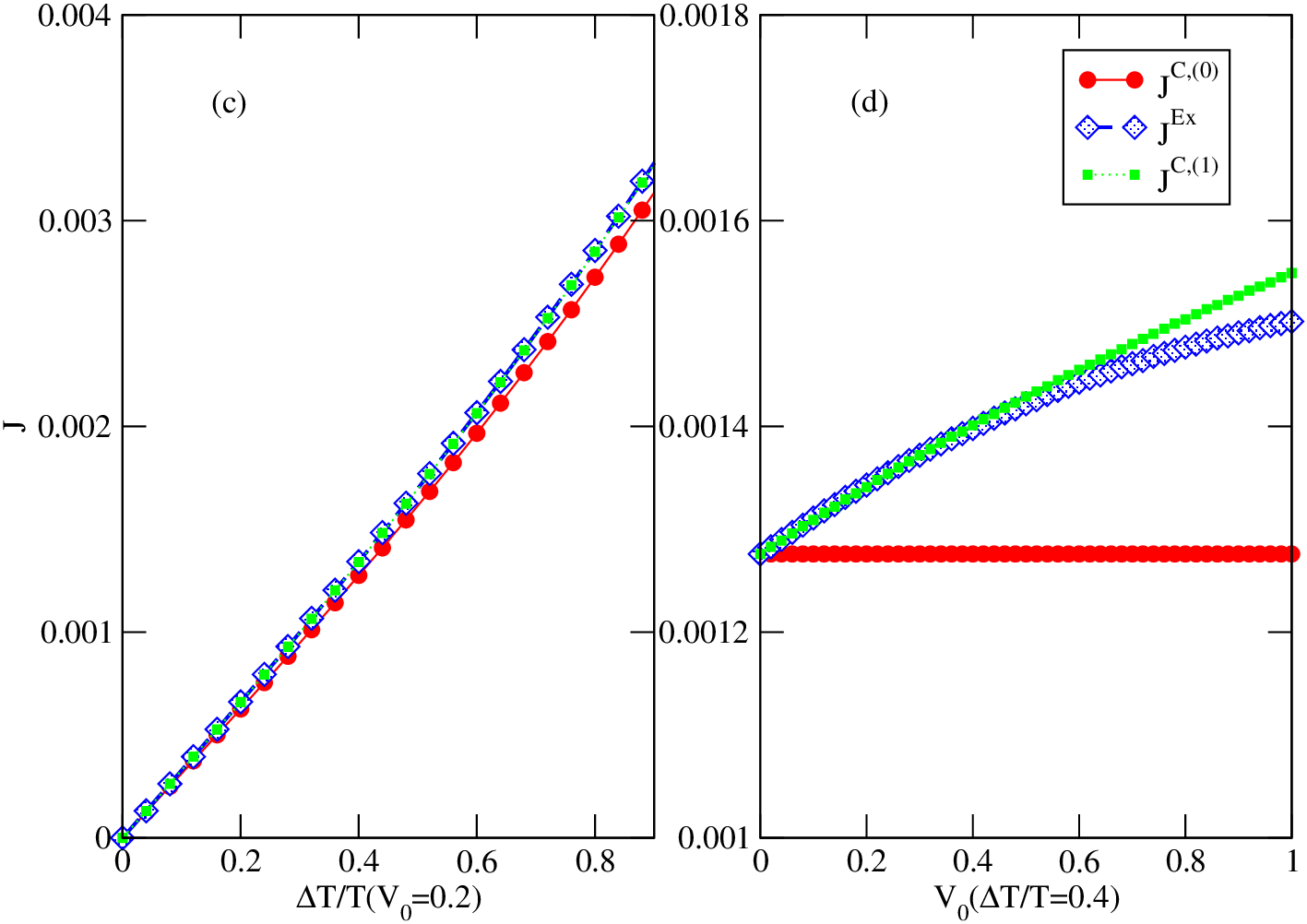}
\caption{\label{fig:cluster} $g^{C, (1)}_{1}$ is compared against $g^{Ex}_{1}$ for non-equilibrium interacting systems. (a) $V=0.2$, accuracy was checked with different values of $\Delta T$. (b) $\Delta T=0.4T$, accuracy was checked with different values of $V_{0}$. In both cases, $d^{C, (1)}$ is always much smaller than $d^{C, (0)}$. $J^{C, (0)}$ and $J^{C, (1)}$ are compared against $J^{Ex}$ in (c) and (d). From (c) we see that for a given value of $V_{0}$, as long as $V_{0}$ is not too large, $J^{C, (0)}$ already provides a major part. From (d) we find that for relatively larger $V_{0}$, the difference between $J^{C, (1)}$ and $J^{C, (0)}$ becomes more important. At the same time, the difference between $J^{C, (1)}$ and $J^{Ex}$ also becomes larger for larger $V_{0}$.}
\end{figure}

Similarly we define $\Delta^{C, (0)}_{1}$ ($d^{C, (0)}$) as the absolute (relative) distance between $g^{C, (0)}_{1}$ and $g^{Ex}_{1}$, and $\Delta^{C, (1)}_{1}$ ($d^{C, (1)}$) as the absolute (relative) distance between $g^{C, (1)}_{1}$ and $g^{Ex}_{1}$. 
First, we set $V_{0}=0.2$ as a constant, and check the accuracy of $g^{C, (1)}_{1}$ with different values of $\Delta T$. From Fig.\ref{fig:cluster}(a) we can see that the worst case is about $d^{(1)}=1\%$. Secondly, we set $\Delta T=0.4T$ as a constant, and check the accuracy of $g^{C, (1)}_{1}$ with different values of $V_{0}$. The worst case is about $d^{(1)}=2\%$ as shown in Fig.\ref{fig:cluster}(b). Overall, $d^{C, (1)}$, is always much smaller than $d^{C, (0)}$.  
From Fig.\ref{fig:cluster}(c) we see that for a small $V_{0}$, $J^{C, (0)}$ already provides a major part. However, in Fig.\ref{fig:cluster}(d) when $V_{0}$ becomes larger, the difference between $J^{C, (1)}$ and $J^{C, (0)}$ becomes more important. We should also note that for larger $V_{0}$, $J^{C, (1)}$ starts to deviate from $J^{Ex}$. This indicates that the approximation captures the essential part of the interaction but it is more accurate for small $V_{0}$. Furthermore, it is likely the approximation could be improved: by keeping $\mathfrak{G}_{2}$ but ignoring correlations in higher order GFs and calculating perturbatively further terms in $V_{0}$ in operators $\hat{m}$. 

In order to estimate the accuracy of this approximation, let us assume $\lambda^2$ and $V_{0}$ are small. Define similarly $\Delta^{C, (0)}_{n}=g^{C, (0)}_{n}-g^{Ex}_{n}$ and $\Delta^{C, (1)}_{n}=g^{C, (1)}_{n}-g^{Ex}_{n}$. Again we start from the equations of the three: $g^{C, (0)}_{1}$, $g^{C, (1)}_{1}$ and $g^{Ex}_{1}$, and then compare the three equations while ignoring certain higher-order terms such as terms which are proportional to $\lambda^2V_{0}$. See Appendix \ref{sec:estimation} for detail of those equations and the estimation. We arrived at, 
\begin{align}
\Delta^{C, (0)}_{1} =  -iV_{0}\left(\Gamma^{(1)}_{0}\right)^{-1}g^{Ex}_{2} \sim V_{0}\left|g^{Ex}_{1}\right|^2,
\end{align}
and
\begin{widetext}
\begin{align}
\Delta^{C, (1)}_{1} = \left(\Gamma^{(1)}_{0}\right)^{-1}\left[iV^{2}_{0}\left(\Gamma^{(2)}_{0}\right)^{-1} g^{Ex}_{3} + \lambda^2V_{0}\left(\Gamma^{(2)}_{0}\right)^{-1}\Delta^{C, (0)}_{1}\right] \sim V^{2}_{0}\left|g^{Ex}_{1}\right|^3 + \lambda^2V^{2}_{0}\left|g^{Ex}_{1}\right|^2 .
\EqLabel{Cluster:DeltaFinal}
\end{align} 
\end{widetext}
This agrees with the numerical tests that $\Delta^{C, (0)}_{1}$ is proportional to $V_{0}$ while $\Delta^{C, (1)}_{1}$ is proportional to $V^{2}_{0}$. We refer readers to Appendix \ref{sec:estimation} for definitions of all $\Gamma$ matrices. Most importantly, we see again that $\Delta^{C, (0)}_{1}$ is multiplied by a small number $\lambda^2V_{0}$ and then becomes a part of $\Delta^{C, (1)}_{1}$. The other term, $V^{2}_{0}g^{Ex}_{3} \sim V^{2}_{0}\left|g^{Ex}_{1}\right|^3$, since roughly $\left|g^{Ex}_{n}\right|=\left|g^{Ex}\right|^{n}$, is also much smaller than $\Delta^{C, (0)}_{1} \sim V_{0}\left|g^{Ex}_{1}\right|^2$. However, for large enough $V_{0}$ the other approximation used in this method, the perturbation expansion of operators $\hat{m}$, will be invalid. Therefore, as long as $V_{0}$ is a small number compared with $t$ then the method under consideration is very reasonable. It should be noted that this method is capable of dealing with large systems since it does not involve a direct diagonalization of a $2^N$-dimension matrix. Although in above examples in order to be comparable with exact solution, only $N=4$ is used, on a normal PC a system $N=100$ has been successfully tested.
 
\section{Conclusion and Discussion}
To conclude, a BBGKY-like equation hierarchy is derived from the Redfield equation and two systematic approximations are suggested to solve the hierarchy. Using the first-order form of the two methods, non-equilibrium stationary states of interacting systems are calculated. It is found that they are consistent with results made available by other direct methods. We also estimate accuracy of the two approximations. The difference among the two is also discussed that the first method is applicable for large $V_{0}$ while the second method is more efficient so it can be applied to larger systems. We have not tested performance of further orders of both methods, although it seems quite straightforward. Our method can also be applied to the local-operator Lindblad equation\cite{Lindblad, Michel}. It may also be worth pursuing a comparison between results from the Lindblad equation and from the Redfield equation. Besides its application on non-equilibrium stationary states, these methods may also be valuable for perturbation theory on equilibrium states. There, using the second method, the equilibrium interacting $g^{C, (1)}_{1}$ can be calculated starting from the equilibrium non-interacting $g^{C, (0)}_{1}$ by setting $T_{L}=T_{R}$ and $\mu_{L}=\mu_{R}$. The computational cost, being a sequence of linear systems with dimension $N^2$, is obviously cheaper than direct diagonalization. It will be interesting to see a further comparison of accuracy and efficiency between these methods and other perturbative methods on equilibrium states. The non-equilibrium equal-time GFs calculated by the proposed methods are also objects of the NEGF method. Investigating relations between these methods and the NEGF method may also be interesting. 

\appendix
\section{Perturbation decomposition of $\hat{m}$s}
\label{sec:mPerturbation}
From its definition in Eq\eqref{ms}, in the representation of eigenvalues $\left\{E_{m}\right\}$ and eigenstates $\left\{|m\rangle \right\}$ of $H_{S}$, operator $\hat{m}$ can be written as
\begin{widetext}
\begin{align}
(\hat{m}_{\alpha})_{mn} = \left(c_{\alpha}\right)_{mn}\sum_{k}|V^{\alpha}_{k}|^2\int_{0}^{\infty}d\tau e^{i\left(E_{n}-E_{m}-\omega_{k,\alpha}\right)\tau} \langle 1-n_{\alpha}\left(\omega_{k,\alpha}\right)\rangle   \notag \\
= \left(c_{\alpha}\right)_{mn}\pi\int d\omega D^{\alpha}\left(\omega\right)|V^{\alpha}\left(\omega\right)|^2 \langle 1-n_{\alpha}\left(\omega\right)\rangle  \delta\left(\Omega_{mn}-\omega\right) \notag \\
= \left(c_{\alpha}\right)_{mn}\pi D^{\alpha}\left(E_{n}-E_{m}\right)|V^{\alpha}\left(E_{n}-E_{m}\right)|^2 \langle 1-n_{\alpha}\left(E_{n}-E_{m}\right)\rangle  ,
\end{align}
\end{widetext}
where we have used $\int_{0}^{\infty}d\tau e^{i\omega\tau} = \pi \delta\left(\omega\right) + iP\left(\frac{1}{\omega}\right)$ and neglected the principal value part. We have also assumed that it is possible to perform a change of variable on $V^{\alpha}_{k}$ such that it becomes $V^{\alpha}\left(k_{nm}\right)$, where $k_{mn}$ is
defined by $\omega_{k_{mn},\alpha} = \Omega_{mn}$, i.e. a bath mode
resonant with this transition. This limits the possible forms of $V^{\alpha}_{k}$ and $\omega_{k,\alpha}$. For example, for a given energy $\Omega_{mn}$, there should be a unique value of $V^{\alpha}_{k_{nm}}$. In this work, we take $V^{\alpha}_{k}$ as a constant so this condition is satisfied. $D_\alpha(\omega)$ is the bath's density of states. We arrived at 
\begin{widetext}
\begin{subequations}
\label{Exms}
\begin{align}
\hat{m}_\alpha = \pi \sum_{m,n}^{} |m\rangle\langle n|  
\langle m|c_{\alpha}|n\rangle 
\left(1-n_\alpha\left(\Omega_{nm}\right)\right)
D_\alpha\left(\Omega_{nm}\right)|V^{\alpha}_{k_{nm}}|^2, \\
\hat{\bar{m}}_\alpha = \pi \sum_{m,n}^{} |m\rangle\langle n|
\langle m|c^{\dag}_{\alpha}|n\rangle 
n_\alpha\left(\Omega_{mn}\right)
D_\alpha\left(\Omega_{mn}\right)|V^{\alpha}_{k_{mn}}|^2,
\end{align} 
\end{subequations}
\end{widetext}
where $\Omega_{mn} = E_{m}-E_{n}=-\Omega_{nm} $. We furthermore set $V^{\alpha}_{k_{nm}}D_\alpha(\omega)$ as a constant and absorb it into $\lambda^{2}$. This procedure involves a direct diagonalization of the isolated system $H_{S}$. One can avoid this by finding such operators $\hat{m}$ perturbatively.

Next assuming $V_{0}$ is small, we want to express operator $\hat{m}_{\alpha}$ in terms of $\left\{c_{m}\right\}$ and $V_{0}$. When $V_{0}=0$ the system is a tight-binding open chain, the following basis transformation 
\begin{align}
\EqLabel{FFT}
c_{k}=\frac{1}{\sqrt{N}}\sum_{l=1}^{N}\sin{\frac{kl\pi}{N+1}}c_{l},
\end{align}
diagonalizes $H_{0}$,
\begin{align}
\EqLabel{FFTH0}
H_{0}=\sum_{k=1}^{N}\epsilon_{k}c^{\dag}_{k}c_{k},
\end{align}
where 
\begin{align}
\EqLabel{dispersion}
\epsilon_{k}=-2t\cos{\frac{\pi k}{N+1}}.
\end{align}
Therefore, $c_{\alpha}\left(t\right)$ is a linear function of all $c_{m}$,
\begin{align}
c^{(0)}_{l}\left(t\right)=\frac{2}{N+1}\sum_{km}\sin{\frac{\pi k l}{N+1}}\sin{\frac{\pi km}{N+1}}e^{-i\epsilon_{k}t}c_{m}.
\end{align}  
Hence $\hat{m}_{\alpha}$ is also a linear combination of all $c_{m}$s. One can imagine that for small $V_{0}$, $\hat{m}_{L}$ should not be too far from a linear combination. Denoted $c_{l}\left(t\right)$ when $V_{0}=0$ as $c_{l}^{(0)}\left(t\right)$. Starting from treating this as the zeroth order to the full dynamical $c_{l}\left(t\right)$, and expanding 
\begin{align}
c_{l}\left(t\right)=\sum_{n}V^{n}_{0}c_{l}^{(n)}\left(t\right),
\end{align}
we may derive a perturbative equation of $c^{(n)}_{l}\left(t\right)$,
\begin{align}
\dot{c}^{(n)}_{l}= it \left(c^{(n)}_{l-1} + c^{(n)}_{l+1}\right) - id^{(n-1)}_{l},
\EqLabel{Cone}
\end{align}
where the short-hand notation, for non-negative integers $n, n_{1}, n_{2}, n_{3}$,   
\begin{align}
d^{(n)}_{l}=\sum_{\substack{
n_1,n_2, n_3 \\ \sum_{i}n_{i}=n}}\left\{c^{(n_1)}_{l}c^{\dag, (n_2)}_{l-1}c^{(n_3)}_{l-1} + c^{(n_{1})}_{l}c^{\dag, (n_{2})}_{l+1}c^{(n_{3})}_{l+1}\right\}.
\end{align}
Then solution of the above equation can be written as
\begin{widetext}
\begin{align}
c^{(n)}_{l}\left(t\right) = -i\int_{0}^{t}d\tau \frac{2}{N+1}\sum_{km}\sin{\frac{\pi k l}{N+1}}\sin{\frac{\pi k m}{N+1}}e^{-i\epsilon_{k}\left(t-\tau\right)}d^{(n-1)}_{m}\left(\tau\right).
\end{align}
\end{widetext}
Here the initial condition that $c^{(n)}\left(0\right)=0(\forall n\geq1)$ is used. 
Plugging this general solution into Eq\eqref{ms}, after straightforward but tedious algebra we arrive at the decomposition of $\hat{m}_{\alpha}$ and $\hat{\bar{m}}_{\alpha}$ in Eq\eqref{mdecomposition2} with expansion coefficients defined as following,
\begin{widetext}
\begin{subequations}
\begin{align}
\mathfrak{D}_{\alpha;m}=\pi\frac{2}{N+1}\sum_{k}\sin{\frac{\pi k l_{\alpha}}{N+1}}\sin{\frac{\pi km}{N+1}}\left[1-n\left(\epsilon_{k}, T_{\alpha}\right)\right],
\\
\mathfrak{\bar{D}}_{\alpha;m}=\pi\frac{2}{N+1}\sum_{k}\sin{\frac{\pi k l_{\alpha}}{N+1}}\sin{\frac{\pi km}{N+1}}n\left(\epsilon_{k}, T_{\alpha}\right),
\\
\mathfrak{D}_{\alpha;m_{1}m_{2}m_{3}}=\pi\sum_{k, m, k_{1}, k_{2}, k_{2}}\left(\frac{2}{N+1}\right)^{4}\frac{n\left(T_{\alpha}, \epsilon\left(k\right)\right)-n\left(T_{\alpha}, \epsilon\left(k_{1}\right) + \epsilon\left(k_{3}\right) - \epsilon\left(k_{2}\right)\right)}{\epsilon\left(k_{1}\right) + \epsilon\left(k_{3}\right) - \epsilon\left(k_{2}\right)- \epsilon\left(k\right)} \notag \\
\sin{\frac{k\pi l_{\alpha}}{N+1}}\sin{\frac{k_{1}\pi m_{1}}{N+1}}\sin{\frac{k_{2}\pi m_{2}}{N+1}}\sin{\frac{k_3\pi m_{3}}{N+1}}\sin{\frac{k\pi m}{N+1}}\sin{\frac{k_1\pi m}{N+1}} \notag \\
\left(\sin{\frac{k_2\pi\left(m+1\right)}{N+1}}\sin{\frac{k_3\pi\left(m+1\right)}{N+1}} + \sin{\frac{k_2\pi\left(m-1\right)}{N+1}}\sin{\frac{k_3\pi\left(m-1\right)}{N+1}}\right).
\end{align}
\end{subequations} 
\end{widetext}

\section{Proof of Non-equilibrium Wick Theorem}
\label{sec:wick}
In this section, we will prove when $V_{0}=0$\cite{note1},
\begin{widetext}
\begin{align}
G_{2}\left(k^{\dag}_{1}, k^{\dag}_{2}, k_{3}, k_{4}\right) = G_{1}\left(k_{1}^{\dag}, k_{4}\right)G_{1}\left(k_{2}^{\dag}, k_{3}\right) -G_{1}\left(k_{1}^{\dag}, k_{3}\right)G_{1}\left(k_{2}^{\dag}, k_{4}\right).
\EqLabel{eq:wick}
\end{align}
\end{widetext}
Here working in the momentum representation, defined in equations from Eq\eqref{FFT} to Eq\eqref{dispersion}, is more convenient than the position representation.
Starting from Eq\eqref{eq:A} with $H_{0}$ in momentum space defined in Eq\eqref{FFTH0} and using $A=c^{\dag}_{k_{1}}c_{k_{2}}$ and $A=c^{\dag}_{k_{1}}c^{\dag}_{k_{2}}c_{k_{3}}c_{k_{4}}$, we have the equations of respectively $G_{1}\left(k_{1}^{\dag}, k_{2}\right)$ and $G_{2}\left(k^{\dag}_{1}, k^{\dag}_{2}, k_{3}, k_{4}\right)$ as following
\begin{widetext}
\begin{subequations}
\begin{align}
0=i\left(\epsilon_{k_{2}}-\epsilon_{k_{1}}\right)G_{1}\left(k^{\dag}_{1}, k_{2}\right) -\lambda^2\frac{2\pi}{N+1}\sum_{\alpha}\sin{\frac{k_{1}\pi l_{\alpha}}{N+1}}\sin{\frac{k_{2}\pi l_{\alpha}}{N+1}}\left(n\left(k_{1}\right)+n\left(k_{2}\right)\right) \notag \\
+ \lambda^2\frac{2\pi}{N+1}\sum_{\alpha, k}\left[\sin{\frac{k_{2}\pi l_{\alpha}}{N+1}}\sin{\frac{k\pi l_{\alpha}}{N+1}}G_{1}\left(k^{\dag}_{1}, k\right) + \sin{\frac{k_{1}\pi l_{\alpha}}{N+1}}\sin{\frac{k\pi l_{\alpha}}{N+1}}G_{1}\left(k^{\dag}, k_{2}\right)\right]
\EqLabel{H0G1}
\\
0=i\left(\epsilon_{k_{4}}+\epsilon_{k_{3}}-\epsilon_{k_{2}}-\epsilon_{k_{1}}\right)G_{2}\left(k^{\dag}_{1}, k^{\dag}_{2}, k_{3}, k_{4}\right) \notag \\
+\lambda^2\frac{2\pi}{N+1}\sum_{\alpha, k}\sin{\frac{k_{1}\pi l_{\alpha}}{N+1}}\sin{\frac{k\pi l_{\alpha}}{N+1}}G_{2}\left(k^{\dag}, k^{\dag}_{2}, k_{3}, k_{4}\right) 
+\lambda^2\frac{2\pi}{N+1}\sum_{\alpha, k}\sin{\frac{k_{2}\pi l_{\alpha}}{N+1}}\sin{\frac{k\pi l_{\alpha}}{N+1}}G_{2}\left(k^{\dag}_{1}, k^{\dag}, k_{3}, k_{4}\right) \notag \\
+ \lambda^2\frac{2\pi}{N+1}\sum_{\alpha, k}\sin{\frac{k_{3}\pi l_{\alpha}}{N+1}}\sin{\frac{k\pi l_{\alpha}}{N+1}}G_{2}\left(k^{\dag}_{1}, k^{\dag}_{2}, k, k_{4}\right) 
+ \lambda^2\frac{2\pi}{N+1}\sum_{\alpha, k}\sin{\frac{k_{4}\pi l_{\alpha}}{N+1}}\sin{\frac{k\pi l_{\alpha}}{N+1}}G_{2}\left(k^{\dag}_{1}, k^{\dag}_{2}, k_{3}, k\right) \notag \\
+\lambda^2\frac{2\pi}{N+1}\sum_{\alpha}\sin{\frac{k_{2}\pi l_{\alpha}}{N+1}}\sin{\frac{k_{4}\pi l_{\alpha}}{N+1}}G_{1}\left(k^{\dag}_{1}, k_{3}\right)\left(n\left(k_{2}\right)+ n\left(k_{4}\right)\right) \notag \\
-\lambda^2\frac{2\pi}{N+1}\sum_{\alpha}\sin{\frac{k_{2}\pi l_{\alpha}}{N+1}}\sin{\frac{k_{3}\pi l_{\alpha}}{N+1}}G_{1}\left(k^{\dag}_{1}, k_{4}\right)\left(n\left(k_{2}\right)+ n\left(k_{3}\right)\right) \notag \\
-\lambda^2\frac{2\pi}{N+1}\sum_{\alpha}\sin{\frac{k_{1}\pi l_{\alpha}}{N+1}}\sin{\frac{k_{4}\pi l_{\alpha}}{N+1}}G_{1}\left(k^{\dag}_{2}, k_{3}\right)\left(n\left(k_{1}\right)+ n\left(k_{4}\right)\right) \notag \\
+\lambda^2\frac{2\pi}{N+1}\sum_{\alpha}\sin{\frac{k_{1}\pi l_{\alpha}}{N+1}}\sin{\frac{k_{3}\pi l_{\alpha}}{N+1}}G_{1}\left(k^{\dag}_{2}, k_{4}\right)\left(n\left(k_{1}\right)+ n\left(k_{3}\right)\right) 
\EqLabel{H0G2}
\end{align}
\end{subequations} 
\end{widetext}
The combination of the two is a closed linear equation and has a unique solution. Therefore, we just need to find one solution. We first apply Eq\eqref{eq:wick} to Eq\eqref{H0G2} to expand $G_{2}$ into products of $G_{1}$. It is then easy to prove that the resulting equation is equivalent with Eq\eqref{H0G1}, meaning that a solution of Eq\eqref{H0G1} is also a solution of Eq\eqref{H0G2}. For example, if we collect terms with $G_{1}\left(k^{\dag}_{2}, k_{4}\right)$ together, we will have
\begin{widetext}
\begin{align}
G_{1}\left(k^{\dag}_{2}, k_{4}\right)\left\{i\left(\epsilon\left(k_{3}\right)-\epsilon\left(k_{1}\right)\right)G_{1}\left(k^{\dag}_{1}, k_{3}\right) -\lambda^2\frac{2\pi}{N+1}\sum_{\alpha}\sin{\frac{k_{1}\pi l_{\alpha}}{N+1}}\sin{\frac{k_{3}\pi l_{\alpha}}{N+1}}\left(n\left(k_{1}\right)+n\left(k_{3}\right)\right)\right. \notag \\
\left.+ \lambda^2\frac{2\pi}{N+1}\sum_{\alpha, k}\left[\sin{\frac{k_{3}\pi l_{\alpha}}{N+1}}\sin{\frac{k\pi l_{\alpha}}{N+1}}G_{1}\left(k^{\dag}_{1}, k\right) + \sin{\frac{k_{1}\pi l_{\alpha}}{N+1}}\sin{\frac{k\pi l_{\alpha}}{N+1}}G_{1}\left(k^{\dag}, k_{3}\right)\right]\right\},
\end{align}
\end{widetext}
where the term in bracket is zero according to Eq\eqref{H0G1}. Therefore, solutions from Eq\eqref{H0G1} satisfies also Eq\eqref{H0G2}, as long as Eq\eqref{eq:wick}, the non-equilibrium Wick Theorem holds. Since the uniqueness Eq\eqref{eq:wick} has to be satisfied.   
 
\section{Estimation of convergence}
\label{sec:estimation}

In this section we present our estimation of the leading order of $G_{1}$ such as $\Delta^{E, (1)}_{1}$ and $\Delta^{C, (1)}_{1}$. We will see that $\Delta^{E, (1)}_{1}$ in fact involves $\Delta^{E, (0)}_{2}$, which in turn needs equation of $G_{2}$, the second equation of the hierarchy, derived from using $A=c^{\dag}_{m}c^{\dag}_{n}c_{m^{'}}c_{n^{'}}$. A similar equation is needed for estimation of $\Delta^{C, (1)}_{1}$.

\subsection{On $\Delta^{E, (1)}_{1}$ from method $1$}
\label{sec:estimation1}
After dropping the $g_{D}$ term and the term which is proportional to $\lambda^2\Delta T$, and keeping only up to the linear order of $\Delta T$, $g^{Ex}_{1}, g^{E, (0)}_{1}$ and $g^{E, (1)}_{1}$ respectively satisfy 
\begin{subequations}
\begin{align}
\left(\Gamma^{(1)}_{0}+\Gamma^{(1)}_{, T}\Delta T\right)g^{Ex}_{1} =iV_{0}g^{Ex}_{2}+\lambda^2\nu_{0},
\EqLabel{Eq:Exg1} \\
\Gamma^{(1)}_{0}g^{E, (0)}_{1} =iV_{0}g^{E,(0)}_{2}+\lambda^2\nu_{0},
\EqLabel{Eq:g1Zero} \\
\left(\Gamma^{(1)}_{0}+\Gamma^{(1)}_{, T}\Delta T\right)g^{E, (1)}_{1} =iV_{0}g^{E, (0)}_{2}+\lambda^2\nu_{0},
\EqLabel{Eq:g1One}
\end{align}
\end{subequations}
where $\Gamma^{(1)}_{0} + \Gamma^{(1)}_{, T}\Delta T$ is the zeroth and first order in $\Delta T$ from $\Gamma^{(1)}$ of Eq\eqref{Eq:Gammag1}. $\Gamma^{(1)}_{, T}$ denotes formally a derivative of $T$ on $\Gamma^{(1)}$. To consider $\Delta^{E, (0)}_{1}$, one may use Eq\eqref{Eq:Exg1} and Eq\eqref{Eq:g1Zero},
\begin{align}
\Delta^{E, (0)}_{1} =  \Delta T \left(\Gamma^{(1)}_{0}\right)^{-1}\Gamma^{(1)}_{, T} g^{Ex}_{1} + iV_{0}\left(\Gamma^{(1)}_{0}\right)^{-1}\Delta^{E, (0)}_{2}.
\end{align}
$\Delta^{E, (1)}_{1}$ can be estimated from Eq\eqref{Eq:Exg1} and Eq\eqref{Eq:g1One}, 
\begin{align}
\Delta^{E, (1)}_{1} = iV_{0}\left(\Gamma^{(1)}_{0}+\Gamma^{(1)}_{, T}\Delta T\right)^{-1}\Delta^{E, (0)}_{2},
\EqLabel{Eq:Delta}
\end{align}
where $\Delta^{E, (0)}_{2}$ is required. We find that roughly speaking $\Delta^{E, (1)}_{1}$ takes the second term of $\Delta^{E, (0)}_{1}$ but drops the first term. Therefore, next we only need to show that the second, $iV_{0}\left(\Gamma^{(1)}_{0}\right)^{-1}\Delta^{E, (0)}_{2}$ is much smaller than the first, or equivalently smaller than the whole $\Delta^{E, (0)}_{1}$. 

Estimation of $\Delta^{E, (0)}_{2}$ involves the second equation of the hierarchy, i.e. equation of $G_{2}$, which can be derived from substituting Eq\eqref{mdecomposition1}, the expression of operators $\hat{m}$ into Eq\eqref{A:G2G3},
\begin{widetext}
\begin{subequations}
\EqLabel{Eq:G2G3}
\protect
\begin{align}
\EqLabel{subeq:g2g2}
0= it\langle c^{\dag}_{m}c^{\dag}_{n}c_{m^{'}}c_{n^{'}+1}\rangle  + it\langle c^{\dag}_{m}c^{\dag}_{n}c_{m^{'}}c_{n^{'}-1}\rangle  + it\langle c^{\dag}_{m}c^{\dag}_{n}c_{m^{'}+1}c_{n^{'}}\rangle  + it\langle c^{\dag}_{m}c^{\dag}_{n}c_{m^{'}-1}c_{n^{'}}\rangle  \notag \\
- it\langle c^{\dag}_{m}c^{\dag}_{n-1}c_{m^{'}}c_{n^{'}}\rangle  - it\langle c^{\dag}_{m}c^{\dag}_{n+1}c_{m^{'}}c_{n^{'}}\rangle  - it\langle c^{\dag}_{m-1}c^{\dag}_{n}c_{m^{'}}c_{n^{'}}\rangle   - it\langle c^{\dag}_{m+1}c^{\dag}_{n}c_{m^{'}}c_{n^{'}}\rangle  \notag \\
+\lambda^2\sum_{l, \alpha}\langle  \delta_{n^{'}\alpha}d_{\alpha; l}c^{\dag}_{m}c^{\dag}_{n}c_{m^{'}}c_{l} + \delta_{m^{'}\alpha}d_{\alpha; l}c^{\dag}_{m}c^{\dag}_{n}c_{l}c_{n^{'}} + \delta_{n\alpha}\bar{d}_{\alpha; l}c^{\dag}_{m}c^{\dag}_{l}c_{m^{'}}c_{n^{'}} + \delta_{m\alpha}\bar{d}_{\alpha; l}c^{\dag}_{l}c^{\dag}_{n}c_{m^{'}}c_{n^{'}}\rangle  \notag \\
+\lambda^2\sum_{l, \alpha}\langle \delta_{n\alpha}d^{*}_{\alpha; l}c^{\dag}_{m}c^{\dag}_{l}c_{m^{'}}c_{n^{'}} + \delta_{m\alpha}d^{*}_{\alpha; l}c^{\dag}_{l}c^{\dag}_{n}c_{m^{'}}c_{n^{'}} + \delta_{n^{'}\alpha}\bar{d}^{*}_{\alpha; l}c^{\dag}_{m}c^{\dag}_{n}c_{m^{'}}c_{l} + \delta_{m^{'}\alpha}\bar{d}^{*}_{\alpha; l}c^{\dag}_{m}c^{\dag}_{n}c_{l}c_{n^{'}}\rangle   \notag \\
+ iV_{0}\langle c^{\dag}_{m}c^{\dag}_{n}c_{m^{'}}c_{n^{'}}\rangle \left(\delta_{m^{'}+1, n^{'}}+\delta_{m^{'}-1, n^{'}}-\delta_{m+1, n}-\delta_{m-1, n}\right) \\
\EqLabel{subeq:g2g3}
 - iV_{0}\sum_{l=m\pm1, n\pm1}\langle c^{\dag}_{l}c^{\dag}_{m}c^{\dag}_{n}c_{l}c_{m^{'}}c_{n^{'}}\rangle  + iV_{0}\sum_{l=m^{'}\pm1, n^{'}\pm1}\langle c^{\dag}_{l}c^{\dag}_{m}c^{\dag}_{n}c_{l}c_{m^{'}}c_{n^{'}}\rangle  \\
\EqLabel{subeq:g2g1}
-\lambda^2\sum_{\alpha}\left[\delta_{n\alpha}\bar{d}_{\alpha; m^{'}}\langle c^{\dag}_{m}c_{n^{'}}\rangle  + \delta_{m\alpha}\bar{d}_{\alpha; n^{'}}\langle c^{\dag}_{n}c_{m^{'}}\rangle  + \delta_{m^{'}\alpha}\bar{d}^{*}_{\alpha; n}\langle c^{\dag}_{m}c_{n^{'}}\rangle  + \delta_{n^{'}\alpha}\bar{d}^{*}_{\alpha; m}\langle c^{\dag}_{n}c_{m^{'}}\rangle \right]  \notag \\
+\lambda^2\sum_{\alpha}\left[\delta_{n\alpha}\bar{d}_{\alpha; n^{'}}\langle c^{\dag}_{m}c_{m^{'}}\rangle  + \delta_{m\alpha}\bar{d}_{\alpha; m^{'}}\langle c^{\dag}_{n}c_{n^{'}}\rangle  + \delta_{n^{'}\alpha}\bar{d}^{*}_{\alpha; n}\langle c^{\dag}_{m}c_{m^{'}}\rangle  + \delta_{m^{'}\alpha}\bar{d}^{*}_{\alpha; m}\langle c^{\dag}_{n}c_{n^{'}}\rangle \right]  \\
\EqLabel{subeq:g2d}
-\lambda^2V_{0}\sum_{\alpha}\langle \delta_{m^{'}\alpha}c^{\dag}_{m}c^{\dag}_{n}c_{n^{'}}D_{\alpha} - \delta_{n^{'}\alpha}c^{\dag}_{m}c^{\dag}_{n}c_{m^{'}}D_{\alpha} + \delta_{m\alpha}c^{\dag}_{n}c_{m^{'}}c_{n^{'}}\bar{D}_{\alpha} - \delta_{n\alpha}c^{\dag}_{m}c_{m^{'}}c_{n^{'}}\bar{D}_{\alpha}\rangle  \notag \\
-\lambda^2V_{0}\sum_{\alpha}\langle \delta_{n^{'}\alpha}\bar{D}^{\dag}_{\alpha}c^{\dag}_{m}c^{\dag}_{n}c_{m^{'}} - \delta_{m^{'}\alpha}\bar{D}^{\dag}_{\alpha}c^{\dag}_{m}c^{\dag}_{n}c_{n^{'}} + \delta_{n\alpha}D^{\dag}_{\alpha}c^{\dag}_{m}c_{m^{'}}c_{n^{'}} - \delta_{m\alpha}D^{\dag}_{\alpha}c^{\dag}_{n}c_{m^{'}}c_{n^{'}}\rangle .
\end{align}  
\end{subequations}
\end{widetext} 
which will be denoted in the following compactly as
\begin{align}
\Gamma^{(2)}g^{Ex}_{2} = iV_{0}g^{Ex}_{3}+\lambda^2g^{Ex}_{1}+\lambda^2V_{0}g^{Ex}_{D3},
\EqLabel{Eq:fullExg2}
\end{align}
where vector $g^{Ex}_{2}$ is defined similarly with $g^{Ex}_{1}$ as following,
\begin{widetext}
\begin{align}
g^{Ex}_{2}=\left[G_{2}\left(1^{\dag},1^{\dag},1,1\right), G_{2}\left(1^{\dag},1^{\dag},1,2\right), \cdots, G_{2}\left(N^{\dag},N^{\dag},N,N\right)\right]^{T}.
\end{align}
\end{widetext}
Note that some of the elements of $g^{Ex}_{2}$ are naturally zero but we still include them into this vector. $g^{Ex}_{3}$, $g^{Ex}_{1}$ and $g^{Ex}_{D3}$ comes from ordering respectively Eq\eqref{subeq:g2g3}, Eq\eqref{subeq:g2g1} and Eq\eqref{subeq:g2d} in the same way as $g^{Ex}_{2}$. Matrix $\Gamma^{(2)}$ can be read off from Eq\eqref{subeq:g2g2}, for example assuming $m, n, m^{'}, n^{'}$ are all different and not equal to $1$ or $N$ for simplicity, we have
\begin{align}
\Gamma^{(2)}_{mN^3+nN^2+m^{'}N+n^{'}, mN^3+nN^2+m^{'}N+n^{'}+1} = it.
\end{align}
Since $\lambda^2V_{0}\ll V_{0}$ we drop the last $g^{Ex}_{D3}$ term in the following estimation of the accuracy. Therefore, the exact solution and the equilibrium solution satisfy respectively,
\begin{subequations}
\begin{align}
\left(\Gamma^{(2)}_{0}+\Gamma^{(2)}_{, T}\Delta T\right)g^{Ex}_{2} =iV_{0}g^{Ex}_{3}+\lambda^2g^{Ex}_{1},
\EqLabel{Eq:Exg2}
\\
\Gamma^{(2)}_{0}g^{E, (0)}_{2} =iV_{0}g^{E, (0)}_{3}+\lambda^2g^{E, (0)}_{1},
\EqLabel{Eq:g2Zero}
\end{align}
\end{subequations}
where $\Gamma^{(2)}_{0}$ and $\Gamma^{(2)}_{, T}\Delta T$ stands for the zeroth and first order in $\Delta T$ in $\Gamma^{(2)}$. Now $\Delta^{E, (0)}_{2}$ can be analyzed,
\begin{widetext}
\begin{align}
\Delta^{E, (0)}_{2} =  iV_{0}\left(\Gamma^{(2)}_{0}\right)^{-1}\Delta^{E, (0)}_{3} + \left(\Gamma^{(2)}_{0}\right)^{-1}\Gamma^{(2)}_{, T}\Delta T g^{Ex}_{2} + \left(\Gamma^{(2)}_{0}\right)^{-1}\lambda^2\Delta^{E, (0)}_{1}.
\end{align}
\end{widetext}
Here in fact $\Gamma^{(1)}$ and $\Gamma^{(2)}$ have different dimensions. However, in this estimation of order of magnitudes, we ignore this difference and furthermore the matrices are regarded as constants with order $1$.
For the moment, let us focus on the last term of $\Delta^{E, (0)}_{2}$. Recall that we want to compare $iV_{0}\left(\Gamma^{(1)}_{0}\right)^{-1}\Delta^{E, (0)}_{2}$ against $\Delta^{E, (0)}_{1}$. Focusing only on the last term, we have
\begin{align}
iV_{0}\Delta^{E, (0)}_{2} \approx  iV_{0}\lambda^2\Delta^{E, (0)}_{1},
\end{align}
which is much smaller than $\Delta^{E, (0)}_{1}$ as long as $V_{0}\lambda^2\ll 1$. Effect of the other two terms has been discussed in the main text in $\S$\ref{sec:result1}.

\subsection{On $\Delta^{C, (1)}_{1}$ from method $2$}
\label{sec:estimation2}
In this case, after dropping $\Gamma^{(1)}_{D}$ term and terms which are proportional to $\lambda^2V_{0}$, $g^{Ex}_{1}$, $g^{C, 0}_{1}$ and $g^{C, (1)}_{1}$ respectively satisfy the following equations,
\begin{subequations}
\EqLabel{Cluster:g1}
\begin{align}
\EqLabel{Cluster:g1Ex}
\Gamma^{(1)}_{0}g^{Ex}_{1} = iV_{0}g^{Ex}_{2} + \lambda^2\nu_{0},\\
\EqLabel{Cluster:gcZero}
\Gamma^{(1)}_{0}g^{C, (0)}_{1} = \lambda^2\nu_{0}, \\
\EqLabel{Cluster:gc1}
\Gamma^{(1)}_{0}g^{C, (1)}_{1} = iV_{0}g^{C, (1)}_{2} + \lambda^2\nu_{0}.
\end{align}  
\end{subequations}
From Eq\eqref{Cluster:gcZero} and Eq\eqref{Cluster:g1Ex} one may find 
\begin{align}
\Delta^{C, (0)}_{1} =  -iV_{0}\left(\Gamma^{(1)}_{0}\right)^{-1}g^{Ex}_{2}.
\end{align}
Comparing Eq\eqref{Cluster:gc1} and Eq\eqref{Cluster:g1Ex}, one gets
\begin{align}
\EqLabel{Cluster:Delta}
\Delta^{C, (1)}_{1} =  iV_{0}\left(\Gamma^{(1)}_{0}\right)^{-1}\left(g^{C, (1)}_{2} - g^{Ex}_{2}\right).
\end{align}
Note magnitude of $\Delta^{C, (1)}_{2}=\left(g^{C, (1)}_{2} - g^{Ex}_{2}\right)$ is in fact smaller than magnitude of $\Delta^{C, (0)}_{2}=\left(g^{C, (0)}_{2} - g^{Ex}_{2}\right)$, which involves the second equation of the hierarchy, i.e. equation of $G_{2}$. So we may analyze the later to get an upper bound of the former. In this case, one need to substitute Eq\eqref{mdecomposition2} to Eq\eqref{A:G2G3}. The resulting equation will have the same structure with Eq\eqref{Eq:G2G3} but every $d_{\alpha;l}$ and $\bar{d}_{\alpha;l}$ are replaced respectively by $\mathfrak{D}_{\alpha;m}$ and $\mathfrak{D}_{\alpha;m}$, and a similar substitution on $D_{\alpha}$ and $\bar{D}_{\alpha}$. Ignoring terms which are proportional to $\lambda^2V_{0}$, $g^{Ex}_{2}$ and $g^{C, (0)}_{2}$ are respectively the solutions of 
\begin{subequations}
\EqLabel{Cluster:g2}
\begin{align}
\Gamma^{(2)}_{0}g^{Ex}_{2} =iV_{0}g^{Ex}_{3}+\lambda^2g^{Ex}_{1},
\EqLabel{Cluster:Exg2} \\
\Gamma^{(2)}_{0}g^{C, (0)}_{2} = \lambda^2g^{C, (0)}_{1}.
\EqLabel{Cluster:g2Zero}
\end{align}
\end{subequations}
Compare these two equations, we find out that
\begin{align}
\Delta^{C, (0)}_{2}=-iV_{0}\left(\Gamma^{(2)}_{0}\right)^{-1}g^{Ex}_{3} 
- \lambda^2\left(\Gamma^{(2)}_{0}\right)^{-1}\Delta^{C, (0)}_{1}.
\end{align}
Focusing only on the last term, we have
\begin{align}
iV_{0}\Delta^{C, (0)}_{2} \approx  -iV_{0}\lambda^2\Delta^{C, (0)}_{1},
\end{align}
which is much smaller than $\Delta^{C, (0)}_{1}$ as long as $V_{0}\lambda^2\ll 1$. Effect of the first term has been discussed in the main text in $\S$\ref{sec:result2}.


\begin{thebibliography}{99}
\bibitem{note1} In a general non-equilibrium state, one may have $\langle C_{k_{3}}C_{k_{4}}\rangle \neq0$. In that case, the above Wick theorem should have more general form and thus the cluster expansion will also have a more general form. In this work, our choice of coupling makes such GFs to be zero.
\end{thebibliography}
\end{document}